\documentclass[journal,a4paper,10pt]{IEEEtran}
\pdfoutput=1

\usepackage{cite}
\usepackage[pdftex]{graphicx}
\usepackage[cmex10]{amsmath}
\usepackage{url}
\usepackage[caption=false,font=footnotesize]{subfig}
\usepackage{tabulary}

\usepackage[T1]{fontenc}
\usepackage[utf8]{inputenc}

\usepackage{rmsbrt_math}
\usepackage{rmsbrt_IBC}
\usepackage{rmsbrt_algorithm}
\usepackage{rmsbrt_sets}
\usepackage{rmsbrt_colours}

\newcommand*{\PiS}{\Pi_i\left(\setS\right)}
\newcommand*{\PiSorth}{\Pi_i^\bot \left(\setS\right)}

\newcommand*{\Uiktwo}{\mathbf{U}_{i_k}^{(2)}}
\newcommand*{\uikntwo}{\mathbf{u}_{i_k,n}^{(2)}}

\newcommand*{\Ujltwo}{\mathbf{U}_{j_l}^{(2)}}

\newcommand*{\Viktwo}{\mathbf{V}_{i_k}^{(2)}}
\newcommand*{\vikntwo}{\mathbf{v}_{i_k,n}^{(2)}}

\newcommand*{\Vjltwo}{\mathbf{V}_{j_l}^{(2)}}

\newcommand*{\yikone}{\mathbf{y}_{i_k}^{(1)}}
\newcommand*{\yiknone}{\mathbf{y}_{i_k,n}^{(1)}}
\newcommand*{\yiktwo}{\mathbf{y}_{i_k}^{(2)}}
\newcommand*{\yikntwo}{\mathbf{y}_{i_k,n}^{(2)}}

\newcommand*{\Phiiktwo}{\mathbf{\Phi}_{i_k}^{(2)}}
\newcommand*{\Phiikbartwo}{\bar{\mathbf{\Phi}}_{i_k}^{(2)}}

\newcommand*{\tik}{t_{i_k}}
\newcommand*{\tikbar}{\bar{t}_{i_k}}

\newcommand*{\alphaikone}{\alpha_i^{(1)}}
\newcommand*{\alphaiktwo}{\alpha^{(2)}}

\newcommand*{\rikone}{r_{i_k}^{(1)}}

\newcommand*{\rikbarone}{\bar{r}_{i_k}^{(1)}}
\newcommand*{\rikbartwo}{\bar{r}_{i_k}^{(2)}}

\newcommand*{\rikntwo}{r_{i_k,n}^{(2)}}
\newcommand*{\riknbarone}{\bar{r}_{i_k,n}^{(1)}}
\newcommand*{\riknbartwo}{\bar{r}_{i_k,n}^{(2)}}

\newcommand*{\rhoikone}{\rho_{i_k}^{(1)}}
\newcommand*{\rhoiktwo}{\rho_{i_k}^{(2)}}

\newcommand*{\eikntwo}{e_{i_k,n}^{(2)}}
\newcommand*{\eiknbartwo}{\bar{e}_{i_k,n}^{(2)}}

\newcommand*{\Wiktwo}{\mathbf{W}_{i_k}^{(2)}}
\newcommand*{\wikntwo}{w_{i_k,n}^{(2)}}

\newcommand*{\Wjltwo}{\mathbf{W}_{j_l}^{(2)}}

\begin{document}

\title{Distributed Long-Term Base Station Clustering in \\Cellular Networks using Coalition Formation}

\author{Rasmus~Brandt,~\IEEEmembership{Student~Member,~IEEE,}
        Rami~Mochaourab,~\IEEEmembership{Member,~IEEE,}
        and~Mats~Bengtsson,~\IEEEmembership{Senior~Member,~IEEE}%
\thanks{The authors are with the Department of Signal Processing, ACCESS Linn\ae{}us Centre, School of Electrical Engineering, KTH Royal Institute of Technology, Stockholm, Sweden. E-mails: \texttt{rabr5411@kth.se}, \texttt{ramimo@kth.se}, \texttt{mats.bengtsson@ee.kth.se}.}%
\thanks{Preliminary results have previously been presented in \cite{Brandt2015}.}}

\markboth{Submitted to IEEE Transactions on Signal and Information Processing over Networks}{Brandt \MakeLowercase{\text
it{et al.}}: Distributed Long-Term Base Station Clustering in Cellular Networks using Coalition Formation}

\maketitle

\begin{abstract}
Interference alignment (IA) is a promising technique for interference mitigation in multicell networks due to its ability to completely cancel the intercell interference through linear precoding and receive filtering. In small networks, the amount of required channel state information (CSI) is modest and IA is therefore typically applied jointly over all base stations. In large networks, where the channel coherence time is short in comparison to the time needed to obtain the required CSI, base station clustering must be applied however. We model such clustered multicell networks as a set of coalitions, where CSI acquisition and IA precoding is performed independently within each coalition. We develop a \mbox{long-term} throughput model which includes both CSI acquisition overhead and the level of interference mitigation ability as a function of the coalition structure. Given the throughput model, we formulate a coalitional game where the involved base stations are the rational players. Allowing for individual deviations by the players, we formulate a distributed coalition formation algorithm with low complexity and low communication overhead that leads to an individually stable coalition structure. The dynamic clustering is performed using only long-term CSI, but we also provide a robust short-term precoding algorithm which accounts for the intercoalition interference when spectrum sharing is applied between coalitions. Numerical simulations show that the distributed coalition formation is generally able to reach \mbox{long-term} sum throughputs within 10 \% of the global optimum.
\end{abstract}

\vspace{-2ex}
\section{Introduction} \label{sec:introduction}
Multicell coordinated precoding \cite{OptResAllCoordMultiCellSys} is a promising technique for improving the downlink cell edge throughputs in future 5G \cite{Andrews2014} wireless networks. By exploiting channel state information (CSI) at the transmitters (CSI-T), the cooperating base stations can balance the generated desired signals with the generated interference by means of precoding and power control. An example of coordinated precoding is interference alignment (IA) \cite{Cadambe2008}, where all interference is cancelled using linear techniques. There is a cost associated with acquiring the necessary CSI-T however: for each cell that takes part in the cooperation, the interfering channels to all mobile stations in the other cells in the cooperation must be acquired. For frequency-division duplex systems, which we focus on in this work, the feedback load for obtaining the CSI-T scales quadratically with the number of cooperating cells.\footnote{For time-division duplex systems, the base station clustering is concerned by pilot allocation and pilot contamination instead; see e.g. \cite{Mochaourab2015arxiv}.} This holds both for analog \cite{ElAyach2012} and digital \cite{Krishnamachari2013j} feedback. In large-sized networks, it is therefore not tractable for all cells to cooperate \cite{Lozano2013}. Since strong interferers matter more than weak ones, it is clear that the CSI acquisition overhead can be reduced by neglecting the weak interferers.

Base station clustering \cite{Peters2012,Chen2014,Pantisano2013} is an approach for balancing the interference mitigation ability of large clusters against the correspondingly high CSI acquisition overhead. In this paper, we derive the long-term throughputs of the receivers accounting for both the CSI acquisition overhead and the spectral efficiency. The spectral efficiency component of the model is obtained by assuming that IA is used within each cluster, but that no cooperation takes place between clusters. The model only requires CSI statistics, thus forming a foundation for long-term clustering.

Given the long-term throughput model, we approach the problem of dynamic base station clustering through the perspective of coalitional\footnote{We will use the word \emph{coalition} interchangeably with the word \emph{cluster}.} games \cite{Saad2009a}. We model the cells as rational players in a hedonic coalitional game \cite{Dreze1980}. The utilities of the players in the game are based on the long-term throughput model mentioned earlier, with the addition of a stabilizing history set \cite{Saad2012} and a deviation search budget \cite{Mochaourab2015arxiv}. By allowing for individual deviations, where a player leaves its current coalition to join another coalition, we provide a distributed coalition formation algorithm with low complexity and low communication overhead. The algorithm is shown to reach an individually stable \cite{Bogomolnaia2002} coalition structure, and empirical evidence shows that this is done within just a few deviation searches per player.

Intracoalition IA was assumed in the long-term throughput model for tractability. With a less restrictive approach for the short-term precoding design, improved throughputs can be achieved however. Given a coalition structure from the long-term coalition formation, we therefore formulate a robust short-term precoding algorithm based on a weighted minimum mean squared error (WMMSE) criterion. Compared to the original WMMSE algorithm \cite{Shi2011}, our version achieves robustness against the spatially unknown intercoalition interference by an optimal level of diagonal loading \cite{Cox1987} of the precoders and receive filters, based only on CSI statistics and filter norms. The algorithm is distributed over the cells, but requires some limited message exchange between coalitions.

Performance of the full system is evaluated through numerical simulations. We study the impact of the channel coherence time, the impact of time sharing or spectrum sharing between coalitions, and the sum throughput as a function of the signal-to-noise ratio (SNR). In all considered scenarios, the proposed coalition formation algorithm performs remarkably well compared to the sum throughput optimal coalition structure.

\vspace{-2ex}
\subsection{Related Work}
Existing work on dynamic base station clustering using IA includes \cite{Peters2012}, where groups of cells were separated using time sharing. A CSI acquisition overhead model which is similar, but less refined, than our proposed model was used and several heuristics were provided for the clustering. In \cite{Chen2014}, approximated rate losses were used as fixed edge weights in an interference graph together with intercluster spectrum sharing. By applying graph partitioning algorithms, two heuristics were provided for the clustering. The CSI acquisition overhead was not explicitly accounted for, and it is unclear whether this can be done within the proposed model. A robust precoding method based on the MaxSINR algorithm was also proposed. In \cite{Pantisano2013}, a coalitional game in partition form was used for IA-based femtocell clustering. The CSI acquisition overhead was modelled in terms of transmission power which limited the amount of power left for data transmission. Static base station clustering using IA was explored in \cite{Tresch2009}, where a hexagonal cell setup was studied numerically. In \cite{Park2015arxiv}, a geographic partition of the plane based on second-order Voronoi regions was used for static pair-wise base station clustering.

There is also a large body of literature on base station clustering for joint processing over cells (so called \emph{network MIMO}); see e.g. \cite{Hong2013} and references therein. Under this paradigm, the limiting factor is generally the base station backhaul connection, since user data is shared between all cooperating base stations. This is a different formulation from this paper, where only CSI is shared between base stations. Another difference is that the clustering generally is performed using short-term CSI, thus having a significantly higher complexity than our proposed long-term clustering solution.

\vspace{-2ex}
\subsection{Contributions and Outline}
\begin{itemize}
    \item In Secs.~\ref{sec:system_model} and \ref{sec:throughput_model}, we propose a model for the long-term throughputs under intracoalition IA. We further propose a generalized---compared to \cite{Peters2012} and \cite{Chen2014}---frame structure which allows for temporal resource allocation into two phases: intercoalition time sharing and intercoalition spectrum sharing.
    \item In Sec.~\ref{sec:coalition_formation}, a coalitional game is formulated where the involved cells are modelled as rational players with utilities derived from the long-term throughput model. A low complexity distributed coalition algorithm is provided, which is shown to lead to an individually-stable coalition structure. Due to the utility model, the coalition formation can be performed in the long term.
    \item In Sec.~\ref{sec:precoding}, a distributed short-term precoding algorithm which is robust against spatially unknown intercoalition interference is proposed. The method can be implemented with only low communication overhead between the interfering coalitions.
\end{itemize}
Compared to our previous work in \cite{Brandt2015}, in this paper we have extended the system model to the cellular case with multiple users per base station, we use a generalized frame structure modelling the feedback and data transmission, and we apply a generalized deviation model for the coalition formation. We also detail the robust short-term precoding method and provide a significantly expanded simulation section.

\vspace{-2ex}
\subsection{Notation}
We denote matrices ($\mathbf{A}$) and vectors ($\mathbf{a}$) using bold uppercase and lowercase letters, respectively. The $n$th column of matrix $\mathbf{A}$ is $\mathbf{a}_n$. Hermitian transpose is denoted by $\mathbf{A}^\herm$ and matrix inverse is denoted by $\mathbf{A}^{-1}$. The $d \times d$ identity matrix is denoted $\matI_d$, and the all zeros matrix is denoted $\matO$. The operator $\diag{\mathbf{a}}$ gives a diagonal matrix whose diagonal elements are given by the vector $\mathbf{a}$. Sets are written in calligraphic letters and braces, e.g. $\setI = \{ 1, \ldots, I \}$. The cardinality of the set is $\card{\setI}$. The operations $\setminus, \cup$ are set difference and set union, respectively. We denote the standard circularly-symmetric complex Gaussian distribution with $\mathcal{CN} \left( 0, 1 \right)$. The expectation operator is $\ex{\cdot}$, and we abbreviate ``independent and identically distributed'' as i.i.d and ``almost surely'' as ``a.s.''. The $n,k$ binomial coefficient is $\binom{n}{k}$. We abbreviate the phrase ``without loss of generality'' with ``w.l.o.g'' and the phrase ``if and only if'' with ``iff''. The limiting behaviour of a function is written using big O notation as $\mathcal{O} \left( f(x) \right)$.

\setcounter{figure}{1}
\begin{figure*}[t]
    \centering
    \includegraphics[width=\textwidth]{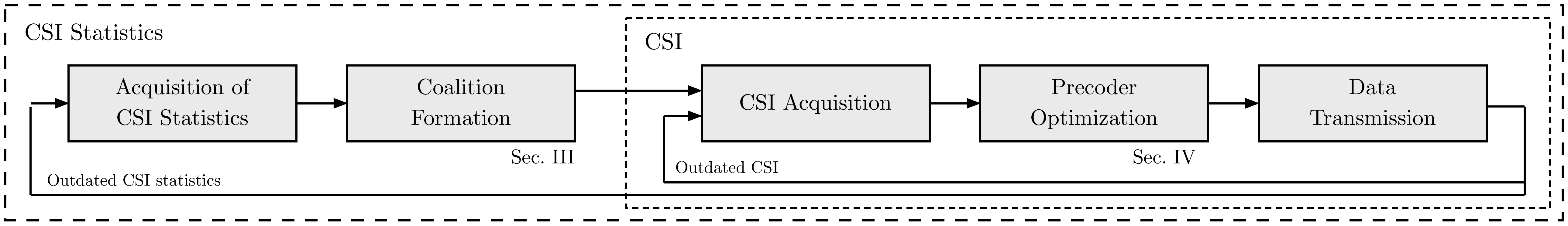}
    \caption{Block diagram of proposed system. Note that the acquisition of the CSI statistics does not necessarily need to be a discrete block, but could be performed concurrently with the other system operation. As a rough approximation for intermediate MS mobility, the CSI statistics must be re-estimated on the order of seconds, whereas the actual CSI must be re-estimated on the order of milliseconds.} \label{fig:block_diagram}
    \vspace{-2ex}
\end{figure*}

\section{System Model} \label{sec:system_model}
We consider a system with $I$ base stations (BSs), indexed using the set $\setI = \{ 1, \ldots, I \}$. Base station~$i$ serves $K_i$ mobile stations (MSs), indexed using the set $\setK_i = \{ 1, \ldots, K_i \}$, with data transmissions in the downlink. Each MS is thus uniquely indexed using the tuple $(i,k)$, which we will often write as $i_k$ for brevity. When a BS  is mentioned together with its associated MSs, we call them a \emph{cell}. BS $i$ has $M_i$ antennas and a total power constraint of $P_i$. MS $i_k$ has $N_{i_k}$ antennas and is served $d_{i_k}$ data streams from BS $i$. We assume that the signal to MS $i_k$ is drawn from a Gaussian codebook, i.i.d. over MSs, such that $\xik \sim \mathcal{CN} \big( \matO, \matI_{d_{i_k}} \big)$. For spatial interference mitigation, we assume that linear processing is applied in the transceivers: BS $i$ uses $\Vik \in \complexnumbers^{M_i \times d_{i_k}}$ as a linear precoder for MS $i_k$, and MS $i_k$ uses $\Uik \in \complexnumbers^{N_{i_k} \times d_{i_k}}$ as a linear receive filter. The wireless channel between BS $j$ and MS $i_k$ is denoted as $\Hikj \in \complexnumbers^{N_{i_k} \times M_i}$. For tractability in Section~\ref{sec:throughput_model}, we assume a simple i.i.d. Rayleigh fading\footnote{This models channels with a large amount of scatterers. It has been shown to, generally, be a simple and good model in urban environments \cite{FundamentalsWirelessCommunication}.} model such that $\left[ \Hikj \right]_{nm} \sim \mathcal{CN} \left( 0, \gamma_{i_kj} \right)$, where $\gamma_{i_kj}$ is determined by the large scale fading.

There is no central controller in the network, and thus all cooperation occurs through peer-to-peer decisions between the BSs. Only BSs that belong to the same cluster can directly cooperate. For consistency with the game theory literature, we refer to a cluster as a \emph{coalition}, and the set of clusters (i.e the base station clustering) as the \emph{coalition structure} \cite{Dreze1980}:
\begin{definition}[Coalition Structure] \label{def:coalition_structure}
    A coalition structure $\setS~=~\{ \setC_1, \setC_2, \ldots, \setC_S \}$ is a partition of $\setI$ into disjoint sets called coalitions, such that $\varnothing \neq \setC_s \subseteq \setI$ for all $\setC_s \in \setS$ and $\bigcup_{p = 1}^S \setC_s = \setI$. For a BS $i \in \setC_s$, we let $\PiS = \setC_s$ map to its coalition and $\PiSorth = \setI \setminus \setC_s$ to the complement.
\end{definition}
\noindent See Fig.~\ref{fig:coalition_outline} for a schematic example of a coalition structure.

\setcounter{figure}{0}
\begin{figure}[t]
    \centering
    \includegraphics[scale=0.8]{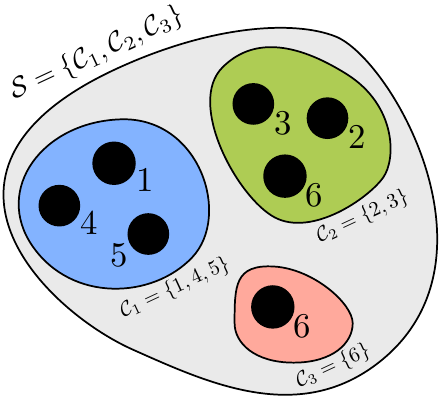}
    \caption{Example of a coalition structure $\setS$ when $\setI = \{ 1, 2, \ldots, 6 \}$. The cells are represented by the black dots, and the coalitions are represented by the coloured areas.} \label{fig:coalition_outline}
    \vspace{-2ex}
\end{figure}
\setcounter{figure}{2}

\subsection{System Operation} \label{sec:system_operation}
We assume that the system operates over two time scales.

\subsubsection{Long Term}
The long-term time scale is determined by the coherence time of the CSI statistics (i.e. $\{ \gamma_{i_kj} \}$), which is dominated by large-scale factors such as path loss and shadowing. We assume that all MSs can estimate their local CSI statistics (i.e. $\{ \gamma_{i_kj} \}_{j \in \setI}$ for MS $i_k$) perfectly without any associated cost, and that the feedback of this information to the serving BS also is cost-free. When the CSI statistics have been acquired at the BSs, coalition formation is performed.

\subsubsection{Short Term}
The short-term time scale is determined by the coherence time of the CSI (i.e. $\{ \Hikj \}$), which is dominated by the small-scale fading. We denote this coherence time as $L_c$, and call this a \emph{coherence block}. In each coherence block, the CSI is perfectly estimated\footnote{The model to be derived can directly be extended to handle imperfect intracluster CSI as well, using the techniques and assumptions in \cite{ElAyach2012,ElAyach2012b}. In order to not detract from the main contribution of this paper, i.e. the throughput model used for the BS clustering, we however omit that direct extension.} at the MSs through pilot transmissions from the BSs such that $\{ \Hikj \}_{j \in \setI}$ is obtained by MS $i_k$. MS $i_k$ then feeds back $\{ \Hikj \}_{j \in \PiS}$ to its serving BS $i$, which in its turn shares the information over a backhaul with the BSs in its coalition $\PiS$. Using the acquired intracoalition CSI-T, precoder optimization is performed independently by the coalitions. A final training stage is then performed, where the MSs estimate their effective channels\footnote{An effective channel is a channel multiplied by a precoder, e.g. $\Hikj \Vjl$.} from all the BSs, thus allowing them to form a receive filter accounting for the effective intercoalition interference.

At the end of a coherence block, the CSI changes and must be estimated again. A schematic of the system operation can be seen in Fig.~\ref{fig:block_diagram}. For clarity, we summarize how the CSI is shared within each coalition:
\begin{definition}[Intracoalition CSI Sharing] \label{def:CSI_sharing}
    The cells in a coalition $\setC_s$ share their local intracoalition CSI such that all cells in $\setC_s$ have access to $\{ \Hikj \}_{i \in \setC_s, k \in \setK_i, j \in \setC_s}$. The cells in $\setC_s$ also have access to the full intercoalition and intracoalition CSI statistics, i.e. $\{ \gamma_{i_kj} \}_{i \in \setC_s, k \in \setK_i, j \in \setI}$.
\end{definition}
We assume that the channel training is performed orthogonally over all cells. This is to ensure that the MSs can estimate their channels with high quality. However, this puts a cap on the size of networks which can be supported; for sufficiently large networks the entire coherence time would be consumed by CSI acquisition (see e.g. \cite{Lozano2013}). For very large scale networks, the techniques proposed herein could however be extended by combining them frequency reuse techniques.

\begin{figure}[t]
    \centering
    \includegraphics[width=\columnwidth]{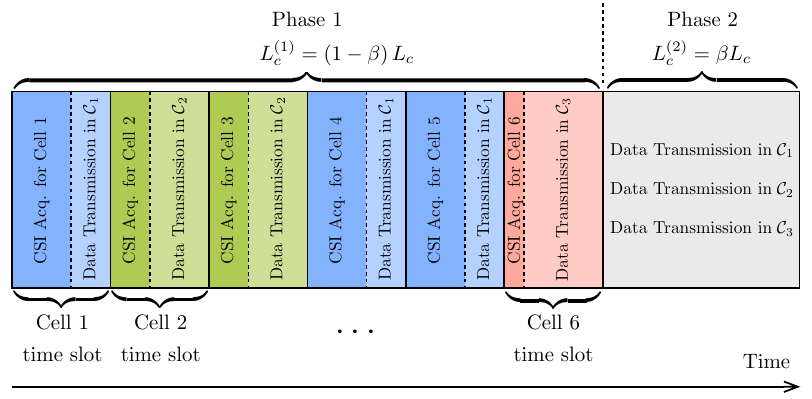}
    \caption{Schematic of proposed frame structure for the coalition structure in Fig.~\ref{fig:coalition_outline}. Note that the total CSI acquisition in phase~1 grows quadratically with the coalition size (cf.~Example~\ref{example:symmetric_CSI_acquisition_analog_feedback}). Full frequency reuse is applied in phase~2.} \label{fig:frame_structure}
    \vspace{-2ex}
\end{figure}

\vspace{-2ex}
\subsection{Management of Intercoalition Interference}
Given a coalition structure, the \emph{intracoalition} interference can be spatially mitigated since the intracoalition CSI is available to all members of the coalition (cf.~Definition~\ref{def:CSI_sharing}). The \emph{intercoalition} interference cannot be spatially mitigated however, due to the lack of intercoalition CSI. We therefore propose two phases for handling this interference.

\subsubsection{Time Sharing}
In phase~1, the coalitions are separated in the time domain (cf.~\cite{Peters2012}), such that no intercoalition interference is received between coalitions. With this approach, the spectral efficiency is high but the temporal resources are not used optimally.

\subsubsection{Spectrum Sharing}
In phase~2, the coalitions are spatially separated (cf.~\cite{Chen2014}), such that only distant BSs contribute unmitigated intercoalition interference. With this approach, the spectral efficiency is lower than in phase~1 due to the unmitigated intercoalition interference, but the temporal resources are maximally used. 

A frame structure parameter\footnote{Due to the lack of a central controller in the network, we let $\beta$ be static and fixed. All cells have a priori knowledge of the value of $\beta$.} $0 \leq \beta < 1$ determines the temporal allocation between the two phases. See Fig.~\ref{fig:frame_structure} for a schematic of the frame structure. phase~1 will generally deliver higher throughputs when the SNR is high and any unmitigated interference is detrimental, whereas phase~2 will generally deliver higher throughputs when the intercoalition interference is negligible compared to the thermal noise. Since the CSI acquisition is performed orthogonally between coalitions, we assume that the phase~1 time allocation is partially used for CSI acquisition, and partially used for data transmission.

\section{Long-Term Throughputs under \\ Intracoalition Interference Alignment} \label{sec:throughput_model}
The total long-term \emph{throughput} of MS $i_k$ is written as a sum of the throughputs in phase~1 and phase~2. In each phase, we model the throughput as a product of a pre-log factor $\alpha$ and a long-term spectral efficiency $\bar{r}$, giving the total throughput as
\begin{align}
    \tikbar & \left( \PiS \right) = \mathbb{E}_{\{ \Hikj \}_{j \in \setI}} \left( \tik(\PiS) \right) \cdot \mathbb{F} \left( \PiS \right) \label{eq:longterm_throughput} \\
    &= \Big( \alphaikone \left( \PiS \right) \rikbarone + \alphaiktwo \rikbartwo \left( \PiS \right) \Big) \cdot \mathbb{F} \left( \PiS \right), \notag
\end{align}
where $\mathbb{F} \left( \PiS \right) = \mathbb{F}_\text{IIA} \left( \PiS \right) \cdot \mathbb{F}_\text{CSI} \left( \PiS \right)$ is a product of feasibility tests, to be defined below.

In the model for the long-term throughput of MS $i_k$ in \eqref{eq:longterm_throughput}, the pre-log factors $\alphaikone \left( \PiS \right)$ and $\alphaiktwo$ describe the temporal degrees of freedom available for data transmission in phase~1 and phase~2, respectively. The long-term spectral efficiencies $\rikbarone$ and $\rikbartwo \left( \PiS \right)$ depend on the signal, interference, and noise powers experienced. In this section we provide a model for how these quantities depend on the coalition structure $\setS$.

In order to model the spectral efficiencies $\rikbarone$ and $\rikbartwo \left( \PiS \right)$, we assume that every coalition performs intracoalition interference alignment (IIA).
\begin{assumption}[Intracoalition Interference Alignment (IIA)] \label{ass:IIA}
    For a coalition $\setC_s \in \setS$, the precoders $\{ \Vik \}_{i \in \setC_s, k \in \setK_i}$ and receive filters $\{ \Uik \}_{i \in \setC_s, k \in \setK_i}$ satisfy
    \begin{align}
        \Uik^\herm \Hikj \Vjl &= \matO, \quad \forall \, j \in \left( \PiS \setminus \{ i \} \right), l \in \setK_i, \label{eq:IA_condition_intercell} \\
        \Uik^\herm \Hiki \mathbf{V}_{i_l} &= \matO, \quad \forall \, l \in \left( \setK_i \setminus \{ k \} \right), \label{eq:IA_condition_intracell} \\
        \hspace{-0.5em} \textnormal{rank} \left( \Uik^\herm \Hiki \Vik \right) &= d_{i_k}. \label{eq:IA_condition_interstream}
    \end{align}
\end{assumption}
Equation \eqref{eq:IA_condition_intercell} ensures that all intracoalition intercell interference is cancelled. Equation \eqref{eq:IA_condition_intracell} ensures that all intracoalition intracell interference is cancelled. Finally, equation \eqref{eq:IA_condition_interstream} ensures that the effective desired channel does not lose rank, thus enabling the transmission of the allocated data streams.

The existence of a solution to \eqref{eq:IA_condition_intercell}--\eqref{eq:IA_condition_interstream} for generic channel matrices---such as in our model in Section~\ref{sec:system_model}---is denoted the \emph{feasibility} of interference alignment. Since the spectral efficiencies $\rikbarone$ and $\rikbartwo \left( \PiS \right)$ will not have any operational meaning unless IIA is feasible, we multiply the throughputs in \eqref{eq:longterm_throughput} with the \emph{IIA feasibility test} as defined by:
\begin{definition}[IIA Feasibility Test] \label{def:IIA_feasibility}
    \begin{equation}
        \mathbb{F}_\textnormal{IIA} \left( \setC_s \right) =
        \begin{cases}
            1 & \text{if IIA is feasible a.s.,} \\
            0 & \text{otherwise.}
        \end{cases}
    \end{equation}
\end{definition}
The feasibility of IA has been thoroughly studied in the literature, see e.g. \cite{Yetis2010,Razaviyayn2012b,Liu2013}, where also several concrete feasibility tests are provided. Note that although we study the case of IA in coalitions, since there is no cooperation between coalitions, the results in \cite{Yetis2010,Razaviyayn2012b,Liu2013} can be applied directly. For a \emph{symmetric}\footnote{In a symmetric network, all BSs serve $K$ MSs with $d$ data streams each. All BSs have $M$ antennas and all MSs have $N$ antennas.} network, an example is given by:
\begin{example}[IIA Feasibility Test in Symmetric Networks] \label{example:IIA_feasibility_in_symmetric_IBC}
    For a coalition $\setC_s \in \setS$ in a symmetric network, a necessary and sufficient condition on a.s. IIA feasibility is \cite[Sec.~V-A]{Liu2013}:
    \begin{equation}
        \mathbb{F}_\textnormal{IIA} \left( \setC_s \right) =
        \begin{cases}
            1 & \textnormal{if} \; \card{\setC_s} \leq \frac{M + N - d}{Kd}, \\
            0 & \textnormal{otherwise}.
        \end{cases}
    \end{equation}
\end{example}
Given an IIA feasible coalition, full intracoalition CSI (as available by Definition~\ref{def:CSI_sharing}) is generally needed in order to find a solution to the conditions in \eqref{eq:IA_condition_intercell}--\eqref{eq:IA_condition_interstream} \cite{Cadambe2008}.

\begin{figure*}
    \begin{equation} \label{eq:phase1_received_signal_with_intracoalition_interference}
        \yikone = \underbrace{\vphantom{\sum_{\substack{j \in \PiS, l \in \setK_j\\(j,l) \neq (i,k)}}}\Uik^\herm \Hiki \Vik \xik}_\text{desired signal} + \underbrace{\sum_{\substack{j \in \PiS, l \in \setK_j\\(j,l) \neq (i,k)}} \hspace{-5mm} \Uik^\herm \Hikj \Vjl \xjl}_\text{intracoalition interference} + \underbrace{\vphantom{\sum_{\substack{j \in \PiS, l \in \setK_j\\(j,l) \neq (i,k)}}}\Uik^\herm \zik}_\text{filtered thermal noise}
    \end{equation}
    \begin{equation} \label{eq:phase2_received_signal_with_intracoalition_interference}
        \yiktwo = \underbrace{\vphantom{\sum_{\substack{j \in \PiS, l \in \setK_j\\(j,l) \neq (i,k)}}}\Uik^\herm \Hiki \Vik \xik}_\text{desired signal} + \underbrace{\sum_{\substack{j \in \PiS, l \in \setK_j\\(j,l) \neq (i,k)}} \hspace{-5mm} \Uik^\herm \Hikj \Vjl \xjl}_\text{intracoalition interference} + \underbrace{\vphantom{\sum_{\substack{j \in \PiS, l \in \setK_j\\(j,l) \neq (i,k)}}}\sum_{j \in \PiSorth, l \in \setK_j} \hspace{-5mm} \Uik^\herm \Hikj \Vjl \xjl}_\text{intercoalition interference} + \underbrace{\vphantom{\sum_{\substack{j \in \PiS, l \in \setK_j\\(j,l) \neq (i,k)}}}\Uik^\herm \zik}_\text{filtered thermal noise}
    \end{equation}
    \hrule
    \vspace{2mm}
    \begin{equation} \label{eq:Lt_example}
        L_t(\setC_s) = \sum_{i \in \setC_s} \big( \hspace{-0.5em} \underbrace{\vphantom{\sum_{j \in \setC_s}}M_i}_{\substack{\text{DL training} \\ \text{for all MSs}}} + \sum_{k \in \setK_i} \big( \hspace{-0.5em} \underbrace{\vphantom{\sum_{j \in \setC_s}}N_{i_k}}_{\substack{\text{UL training} \\ \text{for this MS}}} + \hspace{-0.2em} \underbrace{\vphantom{\sum_{j \in \setC_s}}d_{i_k}}_{\substack{\text{Effective DL training} \\ \text{for this MS}}} + \hspace{-0.5em} \underbrace{\sum_{j \in \setC_s} M_j}_{\substack{\text{Analog feedback} \\ \text{for all intracoalition channels}}} \hspace{-2.5em} \big) \big)
    \end{equation}
    \hrule
\end{figure*}

For tractability in the long-term throughput model, we further make some assumptions on the IIA solution:
\begin{assumption}[Properties of IIA Solution] \label{ass:IA_properties}
    For an IIA solution satisfying Assumption~\ref{ass:IIA}, the following additional properties hold for all $i \in \setI, k \in \setK_i$:
    \begin{enumerate}
        \renewcommand{\theenumi}{\Alph{enumi}}
        \item $\Uik$ is a semi-unitary matrix, i.e. $\Uik^\herm \Uik = \matI_{d_{i_k}}$ \label{ass:IA_properties:bullet:Uik_semiunitary}
        \item $\Vik$ is a matrix with orthogonal columns and uniform power allocation, i.e. $\Vik^\herm \Vik = \frac{P_i}{K_i d_{i_k}} \matI_{d_{i_k}}$ \label{ass:IA_properties:bullet:Vik_semiunitary}
        \item $\uikn$ is statistically independent of $\Hiki \vikn$ and is selected such that $\uikn^\herm \Hiki \vikm = 0$ for all $m \neq n$ \label{ass:IA_properties:bullet:Uik_independence}
        \item $\Vik$ is statistically independent of $\Hiki$ \label{ass:IA_properties:bullet:Vik_independence}
        \item Both $\Uik$ and $\Vik$ are statistically independent of $\Hikj$ for all $j \in \PiSorth$ \label{ass:IA_properties:bullet:joint_Uik_Vik_independence}
    \end{enumerate}
\end{assumption}
The interpretation is that the precoders are used for \mbox{intracell} zero-forcing precoding, and the receive filters are used for intrauser interstream zero-forcing receive filtering. These assumptions are similar to the assumptions in \cite[Lemma~1]{ElAyach2012b}, but generalized to the \emph{cellular} case with \emph{clustered} cells. The existence of a solution to \eqref{eq:IA_condition_intercell}--\eqref{eq:IA_condition_interstream} is not restricted by the assumptions, as given by the following novel result:

\begin{theorem} \label{theorem:ia_solution_assumption_exists}
    If an IIA solution satisfying Assumption~\ref{ass:IIA} exists, there exists an IIA solution satisfying both Assumption~\ref{ass:IIA} and Assumption~\ref{ass:IA_properties}.
\end{theorem}
\begin{IEEEproof}
    The proof is given in Appendix~\ref{proof:ia_solution_assumption_exists}.
\end{IEEEproof}
For the modelling of $\rikbarone$ and $\rikbartwo \left( \PiS \right)$, we now provide a characterization of the effective channels:
\begin{lemma} \label{lemma:effective_desired_channels}
    For an IIA solution satisfying Assumption~\ref{ass:IA_properties}, the effective desired channel for the $n$th stream of MS $i_k$ is
    \begin{equation} \label{eq:fikn_distribution}
        f_{i_k,n} = \uikn^\herm \Hiki \vikn \sim \mathcal{CN} \left( 0, \gamma_{i_ki} \frac{P_i}{K_i d_{i_k}} \right).
    \end{equation}
    The effective intercoalition interfering channel between the $m$th stream intended for MS $j_l$ to the $n$th stream of MS $i_k$, where $j \in \PiSorth$, is
    \begin{equation} \label{eq:gikjlnm_distribution}
        g_{i_kj_l,nm} = \uikn^\herm \Hikj \vjlm  \sim \mathcal{CN} \left( 0, \gamma_{i_kj} \frac{P_j}{K_j d_{j_l}} \right).
    \end{equation}
\end{lemma}
\begin{IEEEproof}
    Due to the bi-unitary invariance of $\Hiki$ \cite{RandomMatrixTheory}, together with Assumptions~\ref{ass:IA_properties}-\ref{ass:IA_properties:bullet:Vik_semiunitary} and \ref{ass:IA_properties}-\ref{ass:IA_properties:bullet:Vik_independence}, each element of $\Hiki \vikn$ is i.i.d. $\mathcal{CN} \left( 0, \gamma_{i_ki} P_i/(K_i d_{i_k}) \right)$. Due to Assumption~\ref{ass:IA_properties}-\ref{ass:IA_properties:bullet:Uik_independence} together with the bi-unitary invariance of $\Hiki \vikn$, the result in \eqref{eq:fikn_distribution} follows. The result for \eqref{eq:gikjlnm_distribution} follows similarly, except that Assumption~\ref{ass:IA_properties}-\ref{ass:IA_properties:bullet:joint_Uik_Vik_independence} is used for the independence.
\end{IEEEproof}

\subsection{Phase~1: Time Sharing for Intercoalition Interference}
In phase~1, the intercoalition interference is handled by time sharing and the intracoalition interference is handled by IIA as per Assumption~\ref{ass:IIA}. The received signal for MS $i_k$ is thus modelled as in \eqref{eq:phase1_received_signal_with_intracoalition_interference}, at the top the page. Note the absence of intercoalition interference due to the time sharing.

\subsubsection{Pre-Log Factor}
The total time allocated to phase~1 is $L_c^{(1)} = (1 - \beta) L_c$. For fairness, we allot $1/I$ fraction of this time to each cell.\footnote{Generalizing to an unequal static allotment over cells is straightforward.} When several cells form a coalition $\setC_s \in \setS$ they each contribute their fraction of time to the coalition, such that the total time available is $\card{\setC_s}/I$. Assuming that the number of symbols needed for CSI acquisition is $L_t \left( \setC_s \right)$, the fraction of time used for CSI acquisition is thus $\frac{L_t \left( \setC_s \right)}{\frac{\card{\setC_s}}{I} L_c^{(1)}}$. For cell $i$, we therefore model the pre-log factor as
\begin{equation} \label{eq:phase1_prelog}
    \begin{split}
        \alphaikone \left( \PiS \right) &= \left( 1 - \beta \right) \frac{\card{\PiS}}{I} \left( 1 - \frac{L_t \left( \PiS \right)}{\frac{\card{\PiS}}{I} L_c^{(1)}} \right) \\
        &= \left( 1 - \beta \right) \left( \frac{\card{\PiS}}{I} - \frac{L_t \left( \PiS \right)}{L_c^{(1)}} \right).
    \end{split}
\end{equation}
The relation in \eqref{eq:phase1_prelog} depends on the specifics of $L_t \left( \setC_s \right)$, which is a function of the CSI acquisition method used. For the case of pilot-assisted channel training and analog feedback, we have the following example:
\begin{example}[CSI Acquisition with Analog Feedback] \label{example:CSI_acquisition_analog_feedback}
    Assume that pilot-assisted channel training and analog feedback is used for the CSI acquisition \cite{ElAyach2012}. Further assume that $M_j~\geq~N_{i_k}$ for all $j \in \setI, i \in \setI, k \in \setK_i$. A model for the \emph{minimum} number of symbols needed for CSI acquisition \cite{Mochaourab2015} in a coalition $\setC_s \in \setS$ is then given by \eqref{eq:Lt_example}, at the top of the page.
\end{example}
In Example~\ref{example:CSI_acquisition_analog_feedback}, the term corresponding to the analog feedback is quadratic in the coalition size $\card{\setC_s}$, thus growing large for large coalitions. The terms corresponding to the channel training are linear in the size of the coalition, in virtue of the broadcast nature of the wireless channel. This model is simplistic, since it assumes \emph{perfect} channel estimation with \emph{minimum} training length.\footnote{Taking the estimation error into account is outside the scope of the current paper, although an interesting venue for future research.} For symmetric networks, Example~\ref{example:CSI_acquisition_analog_feedback} simplifies to the following:
\begin{example}[CSI Acquisition with Analog Feedback in Symmetric Networks]  \label{example:symmetric_CSI_acquisition_analog_feedback}
    For a coalition $\setC_s \in \setS$ in a symmetric network, the \emph{minimum} number of symbols needed for CSI acquisition as given by \eqref{eq:Lt_example} simplifies to
    \begin{equation} \label{eq:Lt_example_symmetric}
        L_t \left( \setC_s \right) = \left( M + K(N + d) \right) \card{\setC_s} + KM \card{\setC_s}^2.
    \end{equation}
\end{example}

Note that larger coalitions benefit by having a large fraction of time available for transmission (cf.~\eqref{eq:phase1_prelog}), but are at a disadvantage since the number of symbols needed for CSI acquisition $L_t \left( \setC_s \right)$ is an increasing function in the coalition size $\card{\setC_s}$. If a coalition were to grow too large, there would not be enough time for the required CSI acquisition. The model would then again (cf.~IIA feasibility) lose its operational meaning, and we therefore multiply the throughputs in \eqref{eq:longterm_throughput} with the \emph{CSI feasibility test} as defined by:
\begin{definition}[CSI Acquisition Feasibility Test]
    \begin{equation}
        \mathbb{F}_\textnormal{CSI} \left( \setC_s \right) =
        \begin{cases}
            1 & \text{if } \frac{\card{\setC_s}}{I} \geq \frac{L_t \left( \setC_s \right)}{L_c^{(1)}}, \\
            0 & \text{otherwise.}
        \end{cases}
    \end{equation}
\end{definition}

\subsubsection{Spectral Efficiency}
Under Assumption~\ref{ass:IA_properties}, the received signal for the $n$th stream of MS $i_k$ in phase~1 (see \eqref{eq:phase1_received_signal_with_intracoalition_interference} and \eqref{eq:fikn_distribution}) can be simplified to
\begin{equation} \label{eq:phase1_received_signal_without_intracoalition_interference}
    \yiknone = f_{i_k,n} \xikn + \uik^\herm \zik.
\end{equation}
The spectral efficiency of \eqref{eq:phase1_received_signal_without_intracoalition_interference} is then given by:
\begin{theorem} \label{theorem:phase1_spectral_efficiency}
    For an IIA solution satisfying Assumption~\ref{ass:IA_properties}, the long-term spectral efficiency in phase~1 for the $n$th stream of MS $i_k$ is
    \begin{equation*}
        \riknbarone = \ex{\log \left( 1 + \abssq{f_{i_k,n}}/\sigma_{i_k}^2 \right)} = e^{1/\rhoikone} E_1 \left( 1/\rhoikone \right),
    \end{equation*}
    where $\rhoikone = \frac{\gamma_{i_ki} P_i/(K_i d_{i_k})}{\sigma_{i_k}^2}$ is the average per-stream SNR and $E_1 \left( \xi \right)~=~\int_{\xi}^\infty t^{-1} e^{-t} \, \mathrm{d}t$ is the exponential integral.\footnote{$E_1(\xi)$ can be calculated numerically, e.g. by summing its truncated power series expansion \cite[5.1.11]{AbramowitzStegun} In Matlab \cite{MATLAB2015a}, it is available as $\texttt{expint} \left( \xi \right)$.}
\end{theorem}
\begin{IEEEproof}
    Given a realization of $f_{i_k,n}$, equation \eqref{eq:phase1_received_signal_without_intracoalition_interference} describes a complex Gaussian channel with spectral efficiency $\log \left( 1 + \abssq{f_{i_k,n}}/\sigma_{i_k}^2 \right)$ \cite[Ch.~9.1]{ElementsInformationTheory}. The result then follows by applying Lemma~\ref{lemma:effective_desired_channels} and performing integration by parts on the expectation integral; see e.g. \cite{Ozarow1994}.
\end{IEEEproof}
Summing up the $d_{i_k}$ streams, we thus write the long-term spectral efficiency for MS $i_k$ in phase~1 as the constant:
\begin{equation} \label{eq:phase1_spectral_efficiency}
    \rikbarone = d_{i_k} e^{1/\rhoikone} E_1 \left( 1/\rhoikone \right).
\end{equation}

\subsection{Phase~2: Spectrum Sharing for Intercoalition Interference}
In phase~2, the whole network operates using spectrum sharing, such that the received signal for MS $i_k$ is given by \eqref{eq:phase2_received_signal_with_intracoalition_interference} at the top of the previous page.

\subsubsection{Pre-Log Factor}
Since all cells share the same time slot in phase~2, the pre-log factor is the constant
\begin{equation}
    \alphaiktwo = \beta.
\end{equation}

\subsubsection{Spectral Efficiency}
Under Assumption~\ref{ass:IA_properties}, the received signal for the $n$th stream of MS $i_k$ in phase~2 (see \eqref{eq:phase2_received_signal_with_intracoalition_interference} and \eqref{eq:gikjlnm_distribution}) can be simplified to
\begin{equation} \label{eq:phase2_received_signal_without_intracoalition_interference}
    \yikntwo = f_{i_k,n} \xikn + \sum_{j \in \PiSorth, l \in \setK_j} \sum_{m=1}^{d_{j_l}} g_{i_kj_l,nm} \xjlm + \uikn^\herm \zik
\end{equation}
The spectral efficiency of \eqref{eq:phase2_received_signal_without_intracoalition_interference} is then given by:
\begin{theorem} \label{theorem:phase2_spectral_efficiency}
    For an IIA solution satisfying Assumption~\ref{ass:IA_properties} and assuming that the intercoalition interference is treated as additional thermal noise in the decoder, the long-term spectral efficiency in phase~2 for the $n$th stream of MS $i_k$ is
    \begin{align*}
        &\riknbartwo \left( \PiS \right) \\
        &= \ex{\log \left( 1 + \frac{\abssq{f_{i_k,n}}}{\sum_{j \in \PiSorth, l \in \setK_j} \sum_{m=1}^{d_{j_l}} \mathbb{E} \abssq{g_{i_kj_l,nm}} + \sigma_{i_k}^2} \right)} \\
        &= e^{1/\rhoiktwo \left( \PiS \right) } E_1 \left( 1/\rhoiktwo \left( \PiS \right)  \right),
    \end{align*}
    where $\rhoiktwo \left( \PiS \right)  = \frac{\gamma_{i_ki} P_i/(K_i d_{i_k})}{\sigma_{i_k}^2 + \sum_{j \in \PiSorth} \gammaikj P_j}$ is the average per-stream signal-to-interference-and-noise ratio (SINR).
\end{theorem}
\begin{IEEEproof}
    Under Definition~\ref{def:CSI_sharing}, the realization of $f_{i_k,n}$ is known to MS $i_k$, but $\{ g_{i_kj_l,nm} \}_{j \in \PiSorth, l \in \setK_j, n = 1, \ldots, d_{i_k}, m = 1, \ldots, d_{j_l}}$ are unknown. Due to the construction of the decoder, the intercoalition interference plays the role of additional Gaussian noise \cite{Lapidoth1996}. This gives the form of the instantaneous spectral efficiency as given inside the expectation operator. The result then follows by applying Lemma~\ref{lemma:effective_desired_channels} and performing integration by parts on the expectation integral.
\end{IEEEproof}
Summing up the $d_{i_k}$ streams, we thus write the long-term spectral efficiency for MS $i_k$ in phase~2 as:
\begin{equation} \label{eq:phase2_spectral_efficiency}
    \rikbartwo \left( \PiS \right) = d_{i_k} e^{1/\rhoiktwo \left( \PiS \right) } E_1 \left( 1/\rhoiktwo \left( \PiS \right)  \right).
\end{equation}

\subsection{Comparison of phase~1 and phase~2} \label{sec:phase_comparison}
The spectral efficiency of phase~1 is noise-limited, since all interference is mitigated through either IIA or time sharing. Due to the time sharing, the temporal resources are not efficiently employed however. Overall, this makes phase~1 suitable for data transmission in high-SNR scenarios, where any unmitigated interference is the main limiting factor in the throughput. The spectral efficiency of phase~2 is interference-limited, since the intercoalition interference is not mitigated. The temporal resources are maximally used however. Overall, this makes phase~2 suitable for intermediate-SNR scenarios, where the CSI acquisition overhead is more important than the unmitigated interference.

The coalition structure $\setS$ affects the pre-log factors of phase~1 (but not the corresponding spectral efficiencies), and the spectral efficiencies of phase~2 (but not the corresponding \mbox{pre-log} factors).

Depending on the scenario, for each MS the data transmission in either phase~1 or phase~2 will be the most efficient. Given a coalition structure $\setS$, the selection of $\beta$ could thus be optimized. Since we are considering a distributed system, we however assume that the $\beta$ is fixed and selected offline.

\section{Low Complexity Long-Term Base Station Clustering through Coalition Formation} \label{sec:coalition_formation}
Given the derived long-term throughput model, it is now our goal to design an algorithm which finds a good coalition structure $\setS$. We consider this problem from the perspective of coalitional games \cite{Saad2009a}, which is a suitable framework for studying the distributed\footnote{Recall that there is no central controller in the network.} formation of coalitions of intelligent and rational \emph{players}, when such formation leads to mutual benefits in terms of the players' \emph{utilities}. In our system, the cells are the players and the corresponding utilities are related to the long-term cell sum throughputs, such that the utility of player~$i$ (i.e. cell~$i$) in the game is defined as
\begin{equation} \label{eq:utility_in_game}
    \tilde{t}_i(\setS;\setH_i,\eta_i) =
        \begin{cases}
            \sum_{k \in \setK_i} \tikbar(\PiS) & \text{if} \; ( \PiS \notin \setH_i \; \text{or} \\
            & \card{\PiS} = 1 ) \; \text{and} \; \eta_i \leq b_i, \\
            0 & \text{otherwise}.
        \end{cases}
\end{equation}
The utility depends on a \emph{history set} $\setH_i \subseteq \{ \setE \in 2^\setI \mid i \in \setE \}$ and a \emph{search budget} $b_i \in \naturalnumbers$, which are introduced for stabilization and complexity reduction, respectively.\footnote{The notion of a history set has previously been used in e.g. \cite{Saad2012}, and a search budget was previously used in e.g. \cite{Mochaourab2015arxiv}.} The history set $\setH_i$ stores the coalitions which player $i$ has been part of before and the quantity $\eta_i \in \naturalnumbers$ denotes the number of times that the player has communicated with other coalitions during the coalition formation. The interpretation of \eqref{eq:utility_in_game} is that the utility of player $i$ is the sum of the long-term throughputs of the MSs in cell $i$, with the restriction that the player never benefits from joining a coalition that it has been a member of before---unless it is the singleton---and the restriction that a player does not benefit if the search budget has been exhausted. Given the utility model, the game associated with our setting is $\langle \setI, \{ \tilde{t}_i \}_{i \in \setI}, \{ b_i \}_{i \in \setI} \rangle$.

The throughputs of the MSs in a particular cell only depend on what other cells participate in the corresponding coalition, and the game is therefore \emph{hedonic} \cite{Dreze1980,Bogomolnaia2002}. A player is not able to share its achieved utility with other players in its coalition, and the utilities are therefore \emph{non-transferable}. Due to the history set and the design of the utilities, i.e. the fact that a player never benefits from joining a coalition that is has been a member of before, the coalition formation algorithm to be proposed will be convergent \cite{Saad2012}.\footnote{For the simpler case of  additively separable utilities, the existence of an individually stable coalition structure for hedonic games--without the introduction of a history set---was shown in \cite{Bogomolnaia2002}.}

For the proposed game we now study \emph{coalition formation}, which is the dynamics that lead to stable coalition structures. We will detail the three main components \cite{Saad2009a} needed in order to describe the coalition formation.

\subsection{Components of Coalition Formation}
\begin{figure}[t]
    \centering
    \includegraphics[scale=0.75]{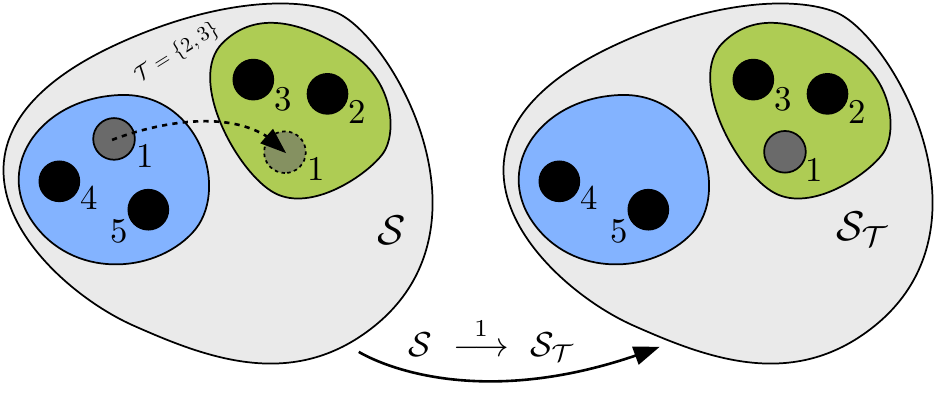}
    \caption{Example of an attach deviation.} \label{fig:deviation1}
\end{figure}

\begin{figure}[t]
    \centering
    \includegraphics[scale=0.75]{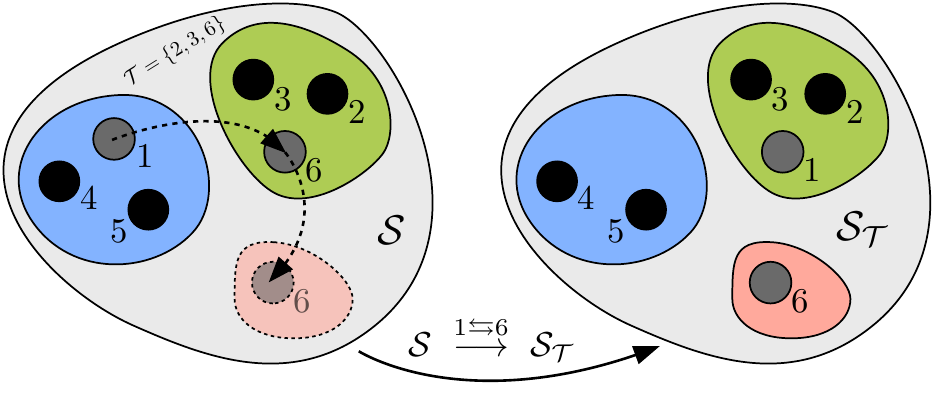}
    \caption{Example of a supplant deviation.} \label{fig:deviation2}
    \vspace{-2ex}
\end{figure}

An individual deviation\footnote{For complexity reasons, we only consider individual deviations where a single player deviates at a time, as inspired by \cite{Dreze1980,Bogomolnaia2002}.} is when a player $i~\in~\setI$ leaves its current coalition $\PiS$ to join another coalition $\setT~\in~\left( \setS \setminus \PiS \right) \cup \{ \varnothing \}$. We propose two types of deviations:
\begin{definition}[Attach Deviation] \label{def:attach_deviation}
    In an \emph{attach deviation}, which we capture with the notation $\setS~\overset{i}{\longrightarrow}~\setS_\setT$, player $i$ simply attaches itself to a coalition $\setT$ such that the coalition structure changes to $\setS_\setT = \left( \setS \setminus \PiS \right) \cup \{ \PiS \setminus \{ i \}, \setT \cup \{ i \} \}$.
\end{definition}
\begin{definition}[Supplant Deviation] \label{def:supplant_deviation}
    In a \emph{supplant deviation}, which we capture with the notation $\setS~\overset{i \leftrightarrows q}{\longrightarrow}~\setS_\setT$, player $i$ supplants another player $q \in \setT$ (the outcast), expelling it to a singleton coalition, such that the coalition structure changes to
    $\setS_\setT = \left( \setS \setminus \PiS \right) \cup \{ \PiS \setminus \{ i \}, \left( \setT \setminus \{ q \} \right) \cup \{ i \}, \{ q \} \}$.
\end{definition}
In the literature, the attach deviation is common (see e.g. \cite{Saad2012,Mochaourab2015arxiv,Dreze1980}). By letting the players either attach, or supplant, (i.e. an \emph{attach-or-supplant} deviation), we allow for a more flexible deviation model, however. This leads to more efficient solutions in some operating regimes (see Section~\ref{sec:performance_evaluation}), than if only the attach deviation was allowed. A visual example of the two deviations are shown in Figs.~\ref{fig:deviation1} and \ref{fig:deviation2}.

\begin{definition}[Admissible Deviation]\label{def:admissible}
    For player $i \in \setI$, a deviation is admissible only if $\tilde{t}_i(\setS_\setT; \setH_i, \eta_i) > \tilde{t}_i(\setS; \setH_i, \eta_i)$.

    An attach deviation $\setS~\overset{i}{\longrightarrow}~\setS_\setT$ is then admissible iff $\tilde{t}_j(\setS_\setT; \setH_j, \eta_j) \geq \tilde{t}_j(\setS; \setH_j, \eta_j), \text{ for all players } j \in \setT$.

    A supplant deviation $\setS~\overset{i \leftrightarrows q}{\longrightarrow}~\setS_\setT$ is then admissible iff $\tilde{t}_j(\setS_\setT; \setH_j, \eta_j) \geq \tilde{t}_j(\setS; \setH_j, \eta_j), \text{ for all players } j \in \setT \setminus \{ q \}$.
\end{definition}

In the attach case, if any existing member of the coalition that is being joined by player $i$ decreases its utility, the player will not be allowed to join. In the supplant case, all members in the coalition being joined, except the outcast $q$, must agree to allow the player $i$ to supplant player $q$. The rational players pursue any admissible deviations, until stability has been reached:
\begin{definition}[Individual Stability]\label{def:Individual_stability}
    A coalition structure $\setS$ is individually stable if there exists no player $i \in \setI$ and coalition structure $\setS_\setT$ such that $\setS~\overset{i}{\longrightarrow}~\setS_\setT$, or $\setS~\overset{i \leftrightarrows q}{\longrightarrow}~\setS_\setT$ for all potential outcasts $q \in \setT \setminus \{ i \}$, are admissible.
\end{definition}
When an individually stable coalition structure has been reached, no player benefits from deviating.


\algrenewcommand\algorithmicindent{1em}
\algtext*{EndIf}
\algtext*{EndLoop}
\begin{algorithm}[t]
    \caption{\label{alg:coalition_formation} Coalition Formation for Base Station Clustering}

    \begin{algorithmic}[1]
    \Statex \textbf{Initialize}: $\setS = \setS^\text{singletons}$, $\setH_i = \varnothing$, $\eta_i = 0$, $\forall \, i \in \setI$
    \Repeat 
    \Loop ~over $i \in \setI$ in lexicographic order
        \Loop{~over $\setT \in \Lambda_i(\setS)$ and $(\setT, q) \in \Delta_i(\setS)$}
        \Statex \hspace{8em} in decreasing order of $\tilde{t}_i(\setS_\setT;\setH_i,\eta_i)$
        \State Increment search factor $\eta_i = \eta_i + 1$
            \If{deviation $\setS~\overset{i \left( \leftrightarrows q \right)}{\longrightarrow}~\setS_\setT$ is admissible}
                \State Replace $\setH_i$ with $\setH_i \cup \PiS$
                \State Replace $\setS$ with $\setS_\setT$
                \State Go to line 2
            \EndIf
        \EndLoop
    \EndLoop
    \Until{no player deviated}
    \end{algorithmic}
\end{algorithm}

\vspace{-2ex}
\subsection{Coalition Formation Algorithm}
Based on the provided concepts in the previous section, we formulate the coalition formation algorithm in Algorithm~\ref{alg:coalition_formation}, on the next page. The initial coalition structure can be arbitrarily selected, but we choose the set of singletons $\setS^\text{singletons}~=~\{ \{ 1 \}, \ldots, \{ I \} \}$ for simplicity. Only a single player deviates at a time, according to the sequence $1,\ldots,I$.

Given its turn, player $i$ forms the set $\setD_i~=~\left( \setS \setminus \PiS \right)~\cup~\{ \varnothing \}$ of all possible coalitions that it could possible deviate to. It further forms the set
\begin{equation*}
    \Lambda_i(\setS) = \{ \setT \in \setD_i \mid \tilde{t}_i(\setS_\setT;\setH_i,\eta_i) > \tilde{t}_i(\setS;\setH_i,\eta_i), \setS~\overset{i}{\longrightarrow}~\setS_\setT \},
\end{equation*}
of all coalitions in attach deviations benefiting itself, and
\begin{align*}
    \Delta_i(\setS) = \{ \left( \setT, q \right) \in \setD_i \times \setI \mid \; &\tilde{t}_i(\setS_\setT;\setH_i,\eta_i) > \tilde{t}_i(\setS;\setH_i,\eta_i), \\
    &q \in \setT, \setS~\overset{i \leftrightarrows q}{\longrightarrow}~\setS_\setT \},
\end{align*}
the set of all coalition-outcast pairs in supplant deviations benefiting itself. The entries of $\Lambda_i(\setS)$ and $\Delta_i(\setS)$ are then sorted in decreasing order of the utility of player $i$, such that the player will try to deviate to the coalitions which benefits it the most first. For a coalition $\setT \in \Lambda_i(\setS)$ that player $i$ wants to attach to, it sends a deviation \emph{proposal} to all members of~$\setT$. Each player in $\setT$ then communicates its \emph{decision} back to player $i$, based on the admissibility criterion in Definition~\ref{def:admissible}. If all members of~$\setT$ allow player $i$ to attach, it stores its old coalition $\PiS$ to the history set and joins $\setT$. For a coalition-outcast pair $(\setT, q) \in \Delta_i(\setS)$, the same protocol can be used as for the attach deviation, except that only the decisions of the non-outcasts $\setT \setminus \{ q \}$ matter (cf.~Definition~\ref{def:admissible}). Note that it is up to the deviating player to propose to treat player $q \in \setT$ as the outcast to the rest of the members of $\setT$. If the outcast $q \in \setT$ is expelled from $\setT$, it receives this message from the other members of $\setT$. For each deviation proposal, player $i$ increments its search count $\eta_i$ by one.

The coalition formation continues until no deviations are admissible. This necessarily eventually happens, as given by \cite[Proof of Thm. 1]{Saad2012}. This result is due to the restriction imposed by the history set: since only a finite number of possible coalition structures exist, and each coalition structure appears at most once in the iterations due to the history set restriction, convergence is guaranteed. Since each player $i$ can only change coalitions a maximum of $b_i$ times, the worst case number of iterations is limited by the by the number of partitions of $\setI$, which is the $I$th Bell number\footnote{The $I$th Bell number describes the number of ways to partition $I$ objects into non-empty disjoint sets \cite[p. 287]{IntroductoryCombinatorics}.}, and $\sum_{i \in \setI} b_i$. Note that the search budgets in themselves guarantee convergence of the algorithm, although we only use them for complexity reduction.

\subsection{Distributed Implementation}
Due to the availability of full local CSI statistics (cf.~Definition~\ref{def:CSI_sharing}), the players can calculate their own utility for all possible deviations. The only message exchange that is needed is to communicate the deviation proposals (from the deviating player to the members of the new coalition) and the deviation decisions (from the members of the new coalition to the deviator and potential outcast).

The history set provides the guarantee of convergence of the algorithm. However, for large $I$, the size of the history set could grow large. For the particular case of bounded coalition sizes (e.g. due to the IIA feasibility in \eqref{eq:Lt_example_symmetric}), the maximum size of the history set is however polynomial in $I$:
\begin{lemma}
    If all formed coalitions $\setC_s$ satisfy $\card{\setC_s} \leq \check{C}$, for some constant $\check{C} \in \naturalnumbers$, then $\card{\setH_i} = \mathcal{O} \left( I^{\check{C}-1} \right)$.
\end{lemma}
\begin{IEEEproof}
    The cardinality of the history set for player $i$ is upper bounded by the total number of coalitions that can form which have player $i$ as a member. Thus,
    \begin{equation}
        \card{\setH_i} \leq \sum_{c = 0}^{\check{C}-1} \binom{I}{c} \leq 1 + \sum_{c = 1}^{\check{C}-1} \frac{I^c}{c!} \leq \check{C} \cdot I^{\check{C}-1},
    \end{equation}
    which gives the result.
\end{IEEEproof}

\begin{figure*}
    \begin{align}
        \rikntwo &= \log \left( 1 + \frac{\abssq{\uikn^\herm \Hiki^\herm \vikn}}{\sum_{\substack{m=1\\m \neq n}}^{d_{i_k}} \abssq{\uikn^\herm \Hiki \vikm} + \sum_{\substack{j \in \setI, l \in \setK_j\\(j,l) \neq (i,k)}} \sum_{m=1}^{d_{j_l}} \abssq{\uikn^\herm \Hikj \vjlm} + \twonormsq{\uikn} \, \sigma_{i_k}^2} \right) \hspace{-0.5em} \label{eq:phase2_WMMSE_spectral_efficiency} \\
        \mathbb{E}_{\{ \Hikj \}_{j \in \PiSorth}} \Big[ \Phiiktwo \Big] &= \sum_{j \in \PiS, l \in \setK_j} \Hikj \Vjltwo \mathbf{V}_{j_l}^{(2),\herm} \Hikj^\herm + \sum_{j \in \PiS, l \in \setK_j} \gammaikj \Fnormsq{\Vjltwo} \matI_{N_{i_k}} + \sigma_{i_k}^2 \matI_{N_{i_k}} = \Phiikbartwo \label{eq:WMMSE_averaged_covariance} \\
        \bar{\mathbf{\Gamma}}_i &= \sum_{j \in \PiS, l \in \setK_j} \Hjli^\herm \Ujltwo \Wjltwo \mathbf{U}_{j_l}^{(2),\herm} \Hjli + \sum_{j \in \PiSorth, l \in \setK_j} \gammajli \tr{\Ujltwo \Wjltwo \mathbf{U}_{j_l}^{(2),\herm}} \, \matI_{M_i} \label{eq:Gamma}
    \end{align}
    \hrule
    \vspace{-4mm}
\end{figure*}

\section{Robust Short-Term Coordinated Precoding through Weighted MMSE Minimization} \label{sec:precoding}
So far we have assumed IIA for the precoding. This allowed for a tractable statistical characterization of the long-term throughputs---which further allowed for the derivation of the distributed coalition formation algorithm---but may not be optimal in terms of spectral efficiency. Since the unaligned intracoalition interference in phase~2 limits the degrees of freedom of the network under IIA precoding, we now propose an alternative coordinated precoding method.\footnote{For the precoding in phase~1, any existing precoding method (see e.g. \cite{Schmidt2013}) can be used since there is no intercoalition interference. In this section we therefore focus on phase~2 precoding, thus denoting the involved quantities with a subscript $(\cdot)^{(2)}$.} By taking into account the statistical information available about the intracoalition interference (cf.~Definition~\ref{def:CSI_sharing}), a robust precoder is designed.

Since we no longer limit ourselves to the requirements of Assumption~\ref{ass:IIA} and Assumption~\ref{ass:IA_properties}, the spectral efficiency of phase~2 is modelled as in \eqref{eq:phase2_WMMSE_spectral_efficiency}, at the top of the next page. For future convenience, we also define the received signal covariance as $\Phiiktwo = \sum_{j \in \setI, l \in \setK_j} \Hikj \Vjltwo \mathbf{V}_{j_l}^{(2),\herm} \Hikj^\herm + \sigma_{i_k}^2 \matI_{N_{i_k}}$.

\subsection{Optimization Problem}
Per Definition~\ref{def:CSI_sharing}, the intracoalition CSI is fully known. The intercoalition CSI is unknown however, and we therefore formulate an optimization problem where the sum throughput performance, averaged w.r.t. to the unknown channels, is maximized:
\begin{align}
    & \underset{\left\{ \uikntwo \right\}, \left\{ \vikntwo \right\}}{\text{maximize}}
    & & \sum_{\substack{i \in \setI, k \in \setK_j\\n = 1,\ldots,d_{i_k}}} \mathbb{E}_{\{ \Hikj \}_{j \in \PiSorth}} \Big[ \rikntwo \Big] \notag \\
    & \text{subject to}
    & & \sum_{k \in \setK_i} \Fnormsq{\Viktwo} \leq P_i, \; i \in \setI. \label{opt:WSRmax}
\end{align}
By applying the procedure pioneered in \cite{Christensen2008} (extended to MIMO in \cite{Shi2011}), we will find a stationary point to a bounded version of \eqref{opt:WSRmax}. The first step is defining the per-stream mean squared error (MSE) as
\begin{align}
    \eikntwo &= \ex{\abssq{\xikn - \mathbf{u}_{i_k,n}^{(2),\herm} \yiktwo}} \\
    &= 1 - 2 \real{\mathbf{u}_{i_k,n}^{(2),\herm} \Hiki \vikntwo} + \mathbf{u}_{i_k,n}^{(2),\herm} \Phiiktwo \uikntwo, \notag
\end{align}
and recalling the following well-known relation between the spectral efficiency and the minimum MSE (MMSE):
\begin{equation} \label{eq:maxrate_maxlogMSE_equivalence}
    \underset{\uikntwo}{\text{max}} \, \rikntwo = \underset{\uikntwo}{\text{max}} \log \left( 1/\eikntwo \right).
\end{equation}
Next, we apply Jensen's inequality to note that
\begin{equation} \label{eq:average_logMSE_bound}
    \mathbb{E}_{\{ \Hikj \}_{j \in \PiSorth}} \Big[ \log \left( \eikntwo \right) \Big] \leq \log \left( \mathbb{E}_{\{ \Hikj \}_{j \in \PiSorth}} \Big[ \eikntwo \Big] \right),
\end{equation}
where the averaged per-stream MSE is
\begin{align}
    &\eiknbartwo = \mathbb{E}_{\{ \Hikj \}_{j \in \PiSorth}} \Big[ \eikntwo \Big] = 1 - 2 \real{\mathbf{u}_{i_k,n}^{(2),\herm} \Hiki \vikntwo} \notag \\
    &+ \mathbf{u}_{i_k,n}^{(2),\herm} \left( \mathbb{E}_{\{ \Hikj \}_{j \in \PiSorth}} \Big[ \Phiiktwo \Big] \right) \uikntwo,
\end{align}
with averaged covariance as in \eqref{eq:WMMSE_averaged_covariance} at the top of the next page.

Applying the relation in \eqref{eq:maxrate_maxlogMSE_equivalence} and the bound in \eqref{eq:average_logMSE_bound} to the optimization problem in \eqref{opt:WSRmax}, and noting that the expected value is finite under our channel model, we write a transformed optimization problem as
\begin{align}
    & \underset{\left\{ \uikntwo \right\}, \left\{ \vikntwo \right\}}{\text{minimize}}
    & & \hspace{-1em} \sum_{\substack{i \in \setI, k \in \setK_j\\n = 1,\ldots,d_{i_k}}} \log \left( \eiknbartwo \right) \notag \\
    & \text{subject to}
    & & \sum_{k \in \setK_i} \Fnormsq{\Viktwo} \leq P_i, \; i \in \setI. \label{opt:transformed_WSRmax}
\end{align}
By assuming that the decoder of MS $i_k$ is oblivious to the statistical distribution of the intercoalition interference and simply treats it as Gaussian noise \cite{Lapidoth1996}, the quantity $-\min_{\uikntwo} \log \left( \eiknbartwo \right)$ in \eqref{opt:transformed_WSRmax} can be interpreted as a spectral efficiency (cf.~Theorem~\ref{theorem:phase2_spectral_efficiency}).

The final optimization problem is now obtained by linearizing the logarithms and including the reciprocals of the linearization points as optimization variables $\{ \wikntwo \}_{n=1}^{d_{i_k}}$ (see e.g. \cite{Shi2011}):
\begin{align}
    & \underset{\substack{\left\{ \uikntwo \right\}, \left\{ \vikntwo \right\}\\ \left\{ \wikntwo \right\}}}{\text{minimize}}
    & & \hspace{-1em} \sum_{\substack{i \in \setI, k \in \setK_j\\n = 1,\ldots,d_{i_k}}} \hspace{-1em} \left( \wikntwo \eiknbartwo - \log \left( \wikntwo \right) - 1 \right) \notag \\
    & \text{subject to}
    & & \sum_{k \in \setK_i} \Fnormsq{\Viktwo} \leq P_i, \; i \in \setI. \label{opt:bounded_linearized_WSRmax}
\end{align}

\subsection{Coordinated Precoding Algorithm}
Applying block coordinate descent to the optimization problem in \eqref{opt:bounded_linearized_WSRmax} results in a distributed and iterative short-term coordinated precoding algorithm. We now briefly summarize these steps. For more details, see e.g. \cite{Shi2011,Komulainen2013}.

The optimality condition for the receive filter of MS $i_k$ is
\begin{equation} \label{eq:optimal_Uik}
    \Uiktwo = \left( \Phiikbartwo \right)^{-1} \Hiki \Viktwo
\end{equation}
with resulting optimal linearization weights
\begin{equation}
    \wikntwo = 1/\eiknbartwo, \quad \forall \, n = 1, \ldots, d_{i_k},
\end{equation}
which for brevity we store in a diagonal matrix
\begin{equation} \label{eq:optimal_Wik}
    \Wiktwo = \diag{\{ \wikntwo \}_{n=1}^{d_{i_k}}} \in \realnumbers^{d_{i_k} \times d_{i_k}}.
\end{equation}
The precoder for MS $i_k$ is then given by
\begin{equation} \label{eq:optimal_Vik}
    \Viktwo = \left( \bar{\mathbf{\Gamma}}_i + \mu_i \, \matI_{M_i} \right)^{-1} \Hiki^\herm \Uiktwo \Wiktwo,
\end{equation}
where $\bar{\mathbf{\Gamma}}_i$ is an virtual uplink covariance matrix as shown in \eqref{eq:Gamma}. The Lagrange multiplier $\mu_i \geq 0$ is found by bisection at BS $i$ such that $\sum_{k \in \setK_i} \Fnormsq{\Viktwo} \leq P_i$.

The full algorithm now consists of consecutively updating the receive filters in \eqref{eq:optimal_Uik}, then updating the linearization weights in \eqref{eq:optimal_Wik}, and then updating the precoders in \eqref{eq:optimal_Vik}. The algorithm is summarized in Algorithm~\ref{alg:robustWMMSE}.

\begin{algorithm}[t]
  \caption{Robust Intracoalition WMMSE Algorithm for Phase~2 Precoding} \label{alg:robustWMMSE}

  \begin{algorithmic}[1]
  \State \textbf{Initialization:} $\{ \Viktwo \}_{i \in \setI, k \in \setK_i}$
  \Repeat
    \State For MS $i_k$, update $\Uiktwo$ in \eqref{eq:optimal_Uik} and $\Wiktwo$ in \eqref{eq:optimal_Wik}
    \State For BS $i$, find $\mu_i \geq 0$ such that $\sum_{k=1}^{K_i} \lvert \lvert \Viktwo \rvert \rvert_\text{F}^2 \leq P_i$
    \State For MS $i_k$, update $\Viktwo$ in \eqref{eq:optimal_Vik}
  \Until{convergence}
  \end{algorithmic}
\end{algorithm}

\begin{theorem}
    Algorithm~\ref{alg:robustWMMSE} converges to a stationary point of the optimization problem in \eqref{opt:transformed_WSRmax}.
\end{theorem}
\begin{IEEEproof}
    This can be shown using the same technique as Theorem~3 of \cite{Shi2011}.
\end{IEEEproof}

\begin{remark}
    When there is no intercoalition interference, the proposed algorithm is identical to the original WMMSE algorithm in \cite{Shi2011}. Thus, Algorithm~\ref{alg:robustWMMSE} could be used for maximizing $\sum_{i \in \setI, k \in \setK_i} \alphaikone \left( \PiS \right) \rikone$ as well.
\end{remark}

\subsection{Distributed Implementation}
In order to form $\Phiikbartwo$ in \eqref{eq:WMMSE_averaged_covariance} and $\bar{\mathbf{\Gamma}}_i$ in \eqref{eq:Gamma}, intracoalition CSI is needed, together with statistical intercoalition CSI. This information is available per Definition~\ref{def:CSI_sharing}. Knowledge of $\{ \lvert \lvert \Vjltwo \rvert \rvert_\text{F}^2 \}_{j \in \PiSorth}$ and $\{ \text{Tr} ( \Ujltwo \Wjltwo \mathbf{U}_{j_l}^{(2),\herm} ) \}_{j \in \PiSorth}$ is also needed. These values must therefore be communicated between coalitions in each iteration, contrary to the assumption of IIA precoding. In order to completely avoid the message exchange between coalitions, the bound in \eqref{eq:average_logMSE_bound} could be further upper bounded by completely disregarding the intercoalition interference. This would completely separate the computation between coalitions, but would also lead to performance degradations.

With the spectral efficiency definition in \eqref{eq:phase2_WMMSE_spectral_efficiency}, the final training stage mentioned in Section~\ref{sec:system_operation} is necessary. Without this final training stage, the rates on the right hand side of \eqref{eq:average_logMSE_bound} would be the achieved rates instead \cite{Lapidoth1996}. We compare these two quantities in the numerical simulation in Sec.~\ref{sec:benchmark_comparison}.

\section{Performance Evaluation} \label{sec:performance_evaluation}
The performance of the proposed system is studied through numerical simulations. The simulation scenario is a large-scale deployment of macro BSs which includes effects like path loss, shadow fading, and small-scale fading. The parameters are generally inspired by 3GPP~Case~2 \cite[A.2.1.1]{TR25814}, except for the small scale fading modelling where we choose i.i.d. Rayleigh fading for simplicity (cf.~Section~\ref{sec:system_model}). We consider a symmetric network where $I = 12$ BSs each serve $K = 2$ MSs with $d = 1$ stream each. The BSs have $M = 8$ antennas each and the MSs have $N = 2$ antennas each. The IIA feasibility test and CSI acquisition overhead are given by Example~\ref{example:IIA_feasibility_in_symmetric_IBC} and Example~\ref{example:symmetric_CSI_acquisition_analog_feedback}, respectively.\footnote{Note that we consider a symmetric network for simplicity, but the proposed algorithms are not limited to symmetric networks in general.} Since base station clustering is more challenging in asymmetric geographies, instead of the typical hexagonal cells, we randomly drop the BSs in a square. The area of the square is selected such that the average cell size is identical to the case of hexagonal cells with an inter site distance of 500 m, as mandated by 3GPP~Case~2. For simplicity, we let the $K$ served MSs be placed uniformly at random at a circle around the serving BS at a distance of $150$ m. We study the performance averaged over $250$ drops of the BSs and MSs, and for each drop we generate $10$ small-scale fading realizations. As given by 3GPP~Case~2, we consider a carrier frequency of $2$ GHz, giving the path loss as $15.3 + 37.6 \log_{10}(\text{distance} \, \text{[m]})$ in decibels. The shadow fading is i.i.d. log-normal with $8$ dB standard deviation. All antennas are omnidirectional with antenna gain $0$ dB. Unless otherwise noted, we let $\beta = 0.5$, $\text{SNR} = \frac{P}{\sigma^2} = 20 \, \text{dB}$ and the MS speed is $30$ km/h. The coherence bandwidth is set as $W_c = 300 \, \text{kHz}$.

We consider the throughputs with \mbox{long-term} spectral efficiency from Theorems~\ref{theorem:phase1_spectral_efficiency} and \ref{theorem:phase2_spectral_efficiency} (``IIA throughputs'') as well as the throughputs with short-term spectral efficiency from \eqref{eq:phase2_WMMSE_spectral_efficiency} (``WMMSE throughputs''). As the figure of merit for the system, we use the sum throughput over all served MSs. For the robust WMMSE algorithm, we perform sufficiently many iterations to reach a relative convergence of $10^{-3}$ of the sum throughput. The initial precoder for MS $i_k$ is selected as the $d_{i_k}$ strongest right singular vectors of $\Hiki$.

For benchmarking of the coalition formation algorithm, we compare to the sum throughput optimal coalition structure as generated by a branch and bound approach \cite{Brandt2016csubmitted}. We also compare to the case of singleton coalitions $\setS^\text{singletons}~=~\{ \{ 1 \}, \ldots, \{ I \} \}$, as well as the grand coalition $\setS~=~\{ \setI \}$. From the literature, we also compare to the \mbox{k-means} clustering of \cite{Chen2014}, whose weight matrix has been extended in the obvious way for the cellular case. This method may give coalitions that are larger than what is strictly IA feasible, and we thus only evaluate its performance together with WMMSE precoding. We also compare to the heuristic grouping method of \cite{Peters2012}, which has been slightly modified to always serve all MSs with equal number of data streams.

\begin{figure}[t]
    \centering
    \includegraphics[scale=0.7]{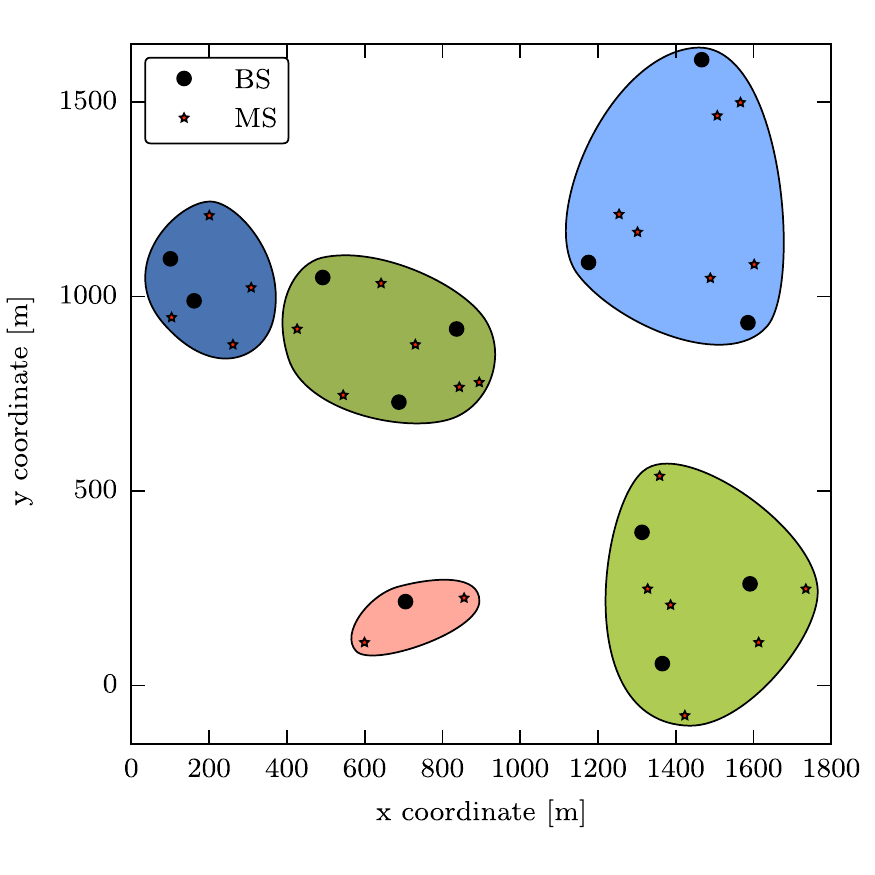}
    \caption{Example of network realization and coalition structure generated by the coalition formation algorithm. The MSs are dropped $150$ m from their serving BS, and with high probability the serving BS is the closest BS.} \label{fig:network_layout_example-annot}
\end{figure}

\begin{figure*}[!t]
    \centering
    \subfloat[Sum throughput under IIA precoding]{\includegraphics{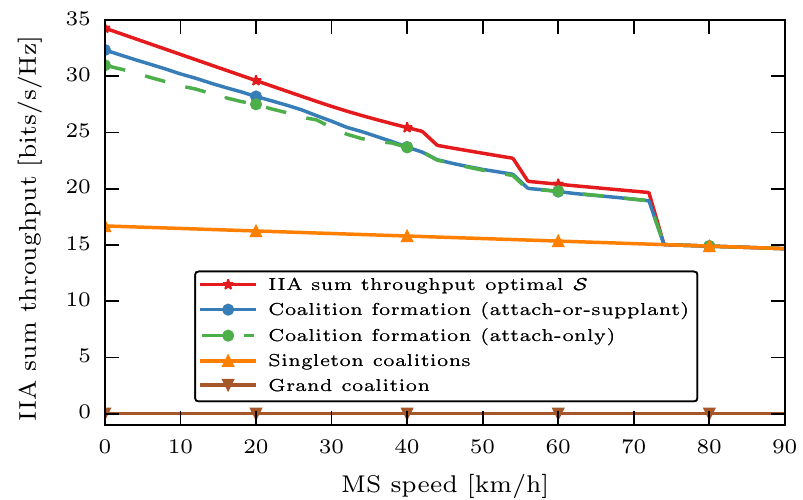} \label{fig:num_coherence_symbols_longterm-sumrate}}
    \hfil
    \subfloat[Sum throughput under WMMSE precoding]{\includegraphics{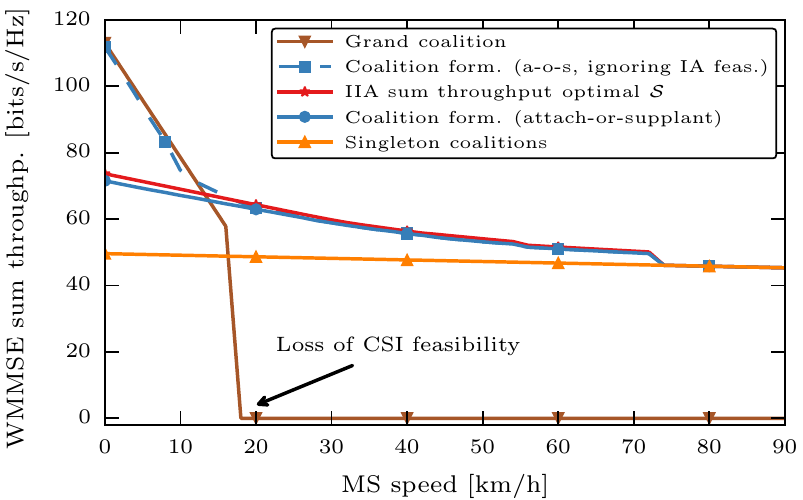} \label{fig:num_coherence_symbols_instantaneous-sumrate}}
    \caption{IIA and WMMSE sum throughput when implicitly varying $L_c$ through the proxy of MS speed.} \label{fig:num_coherence_symbols}
\end{figure*}

\vspace{-3ex}
\subsection{Example Coalition Structure}
First we give an example of a typical network realization, together with the correspondingly generated coalition structure from the coalition formation algorithm with attach-or-supplant deviations; see Fig.~\ref{fig:network_layout_example-annot}. The geographic proximity of cells is important in the distance-dependent path loss, leading to grouping of nearby cells.

\subsection{Coherence Block Length}
\begin{figure}[t]
    \centering
    \includegraphics{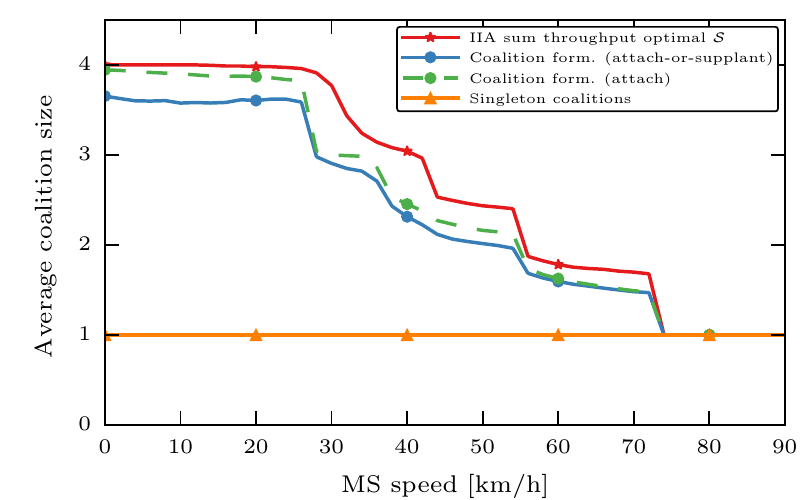}
    \caption{Average coalition size when implicitly varying $L_c$.} \label{fig:num_coherence_symbols_longterm-avg_cluster_size}
    \vspace{-2ex}
\end{figure}

The length of the coherence block $L_c$ directly affects performance in the proposed frame structure and throughput model. In the simulations, we use a simple model (see \cite{Jindal2010} for details) that maps\footnote{This is done by showing that the achievable rate with pilot-assisted training is identical between continuous fading with rectangular Doppler spectrum and block fading if $f_D = 1/(2L_c)$, where $f_D = L_s v/\lambda$ is the Doppler frequency, $L_s$ is the symbol period and $\lambda$ is the carrier wavelength.} the MS speed into an integer $L_c$. Varying the MS speed---thus varying $L_c$---we show the IIA and WMMSE sum throughputs in Fig.~\ref{fig:num_coherence_symbols} (at the top of the next page) and the corresponding average coalition size in Fig.~\ref{fig:num_coherence_symbols_longterm-avg_cluster_size}. As the MS speed increases, $L_c$ decreases, leading to lower performance. The plateauing effects in Fig.~\ref{fig:num_coherence_symbols_longterm-sumrate} and Fig.~\ref{fig:num_coherence_symbols_longterm-avg_cluster_size} are due to the CSI acquisition overhead limiting the size of the coalitions. The coalition formation algorithms are able to track the global optimum, significantly outperforming the case of singleton coalitions. The IIA throughput of the grand coalition is constantly zero, as the grand coalition is not IIA feasible. In the low MS speed regime, the limiting factor of the coalitions is the IIA feasibility and not the CSI acquisition overhead. The coalition formation algorithm with only attach deviations could thus easily get stuck, since there generally is little benefit from going from a larger coalition to a smaller one---an action which might be necessary for efficiency under the hard constraints of IIA feasibility. By allowing for the attach-or-supplant deviation however, more dynamism is allowed, leading to improved sum throughput in the low MS speed regime. In the following, we therefore use the attach-or-supplant version.

The WMMSE sum throughput in Fig.~\ref{fig:num_coherence_symbols_instantaneous-sumrate} shows a substantial improvement over the IIA sum throughput in Fig.~\ref{fig:num_coherence_symbols_longterm-sumrate}, due to the more efficient treatment of the intercoalition interference. Since the robust WMMSE algorithm performs power control, IIA feasibility is not strictly necessary in order to reach good performance. This can be seen for low MS speeds, where the grand coalition is outperforming the methods relying on the long-term IIA throughput model. By heuristically modifying the throughputs used in the coalition formation to ignore the IIA feasibility (``Coalition form. (a-o-s, ignoring IA feas.)'' in the legend), the resulting WMMSE sum throughput follows that of the grand coalition.

\begin{figure}[t]
    \centering
    \includegraphics{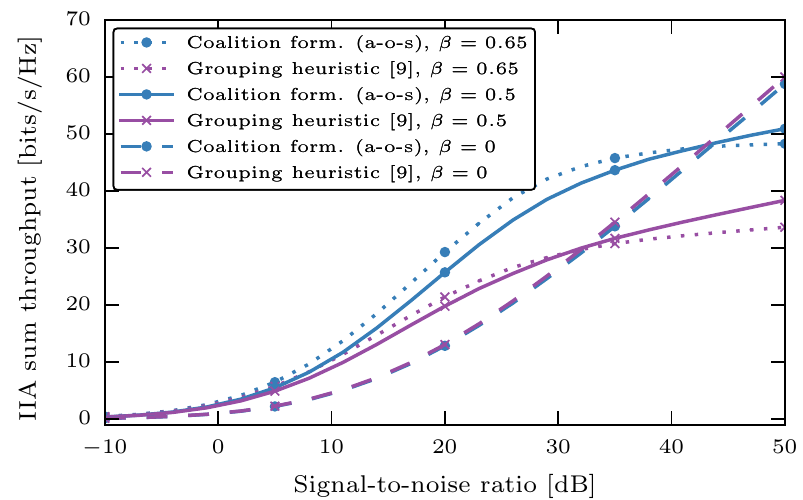}
    \caption{IIA sum throughput for $\beta \in \{ 0, 0.5, 0.65 \}$ and varying SNR.} \label{fig:beta_longterm-sumrate}
    \vspace{-2ex}
\end{figure}

\subsection{Selection of $\beta$}
The frame structure parameter $\beta$ determines temporal resource allocation between phase~1 and phase~2. Depending on the SNR, the sum throughput optimal value of $\beta$ will either be $\beta^{(1), \star} = 0$ (in the high-SNR regime), or close to $\beta^{(2), \star} = \max_{\mathbb{F}_\text{CSI} \left( \setC_s \right) = 1, \forall \setC_s \in \setS} \beta = 0.65$ (in the intermediate SNR regime). To find the boundary between these two regimes, we plot the IIA throughput for the case of $\beta \in \{ 0, 0.5, 0.65 \}$ in Fig.~\ref{fig:beta_longterm-sumrate}. For the proposed coalition formation, the intersection of the three curves is at an SNR of 42 dB. With an SNR lower than 42 dB, it is beneficial to maximize the time in phase~2 (i.e. selecting $\beta = 0.65$). With an SNR higher than 42 dB, it is beneficial to maximize the time in phase~1, (i.e. selecting $\beta = 0$). A good trade-off is achieved by selecting $\beta = 0.5$. The grouping heuristic of~\cite{Peters2012} performs well for $\beta = 0$, which is the setting it was developed for. Its lacklustre performance for $\beta \in \{ 0.5, 0.65 \}$ is due to the fact that only the coalition \emph{sizes} matter in the grouping heuristic, and not the actual members of the coalitions. This leads to poor performance in phase~2.

\setcounter{figure}{10}
\begin{figure*}[!t]
    \centering
    \subfloat[Comparing clustering methods.]{\includegraphics{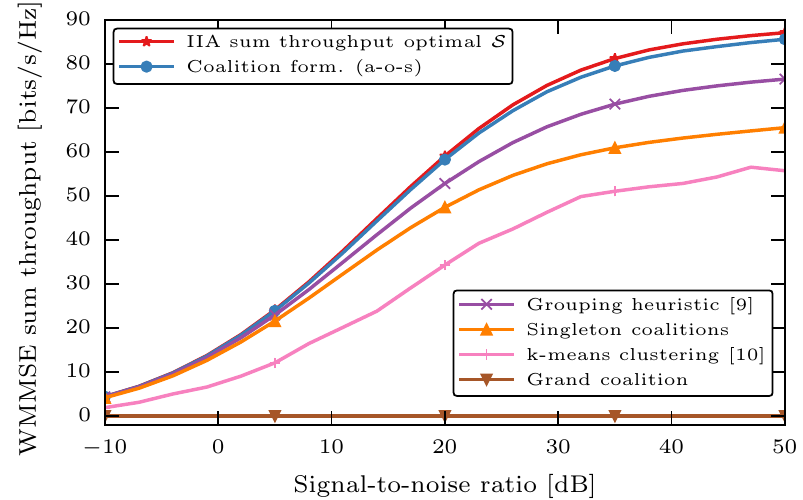} \label{fig:assignment_methods}}
    \hfil
    \subfloat[Comparing precoding methods.]{\includegraphics{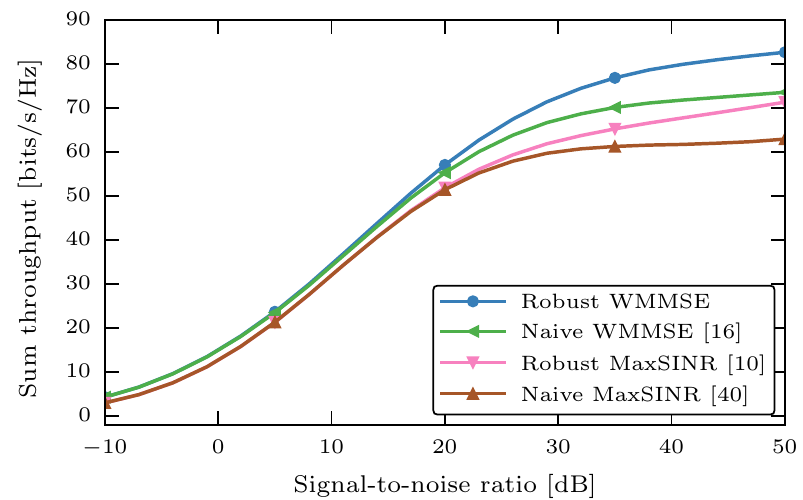} \label{fig:precoding_methods}}
    \caption{Comparison of methods when varying SNR.} \label{fig:methods}
    \vspace{-4mm}
\end{figure*}

\vspace{-2ex}
\subsection{Coalition Formation Complexity for Large Networks}
We now consider a network of $I$ cells randomly placed in a square whose sides give the same average cell size as before. In order to accommodate large networks, we let the MS speed be $3$ km/h. The average number of searches in the coalition formation is given in Fig.~\ref{fig:I_longterm-num_searches}, together with the achieved IIA sum throughput. The coalition formation with attach-or-supplant deviations has a higher complexity than the coalition formation with attach-only deviations. In absolute terms, both algorithms have very low complexities however. The higher complexity of the attach-or-supplant version does pay off in terms of higher achieved IIA sum throughputs.

\setcounter{figure}{9}
\begin{figure}[t]
    \centering
    \includegraphics{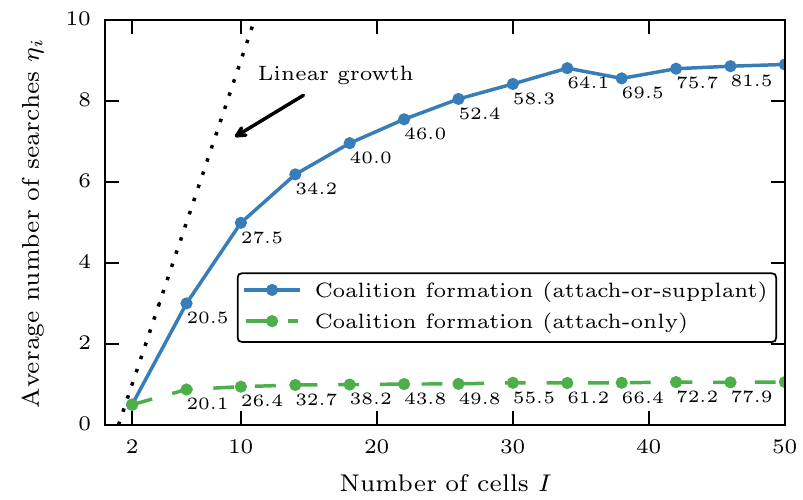}
    \caption{Coalition formation complexity and achieved IIA sum throughput (numbers below curves, [bits/s/Hz]) when varying the number of cells $I$.} \label{fig:I_longterm-num_searches}
    \vspace{-4mm}
\end{figure}

\vspace{-2ex}
\subsection{Comparison with Benchmarks} \label{sec:benchmark_comparison}
In Fig.~\ref{fig:assignment_methods}, we compare the coalition formation algorithm with the benchmarks in terms of WMMSE throughput by varying the SNR. The coalition formation is very close to the WMMSE throughput of the IIA throughput optimal coalition structure, but also outperforms all other benchmarks. The grouping heuristic of \cite{Peters2012}, which is designed for the $\beta = 0$ case, is still better than the singleton coalition structure. Although the \mbox{k-means} algorithm of \cite{Chen2014} is the best algorithm in \cite{Chen2014}, its performance is underwhelming in our context. This may be due to the fact that we are studying the cellular case, as the same implementation of the algorithm has been verified to perform well in the interference channel setting \cite[Fig.~4]{Brandt2015}.

In Fig.~\ref{fig:precoding_methods}, we compare our proposed robust WMMSE algorithm with the original (naive) WMMSE algorithm from \cite{Shi2011}, the robust MaxSINR algorithm from \cite{Chen2014}, and the original (naive) MaxSINR algorithm from \cite{Gomadam2011}. Both robust algorithms are designed to account for unknown intercoalition interference through diagonal loading of the linear filters. The WMMSE approach is superior to the MaxSINR approach in general, and our proposed robust WMMSE algorithm outperforms the benchmarks by between 10––25\% at high SNR.

\vspace{-2ex}
\section{Conclusions} \label{sec:conclusions}
Base station clustering is necessary in large interference alignment-based networks, due to the large CSI acquisition overhead which is otherwise incurred. In this paper, we have used first principles to derive a model for the long-term MS throughputs in such a network. By applying notions from the field of coalitional games, we formulated a distributed coalition formation algorithm which stops when an individually stable coalition structure is reached. The numerical results showed that the distributed algorithm performed close to the optimal solution, with just a few number of iterations. Given intracoalition CSI and intercoalition CSI statistics, a robust WMMSE algorithm was formulated which further improved the sum throughputs. This came at the cost of some limited intercoalition message passing during the WMMSE iterations.

\begin{figure*}
    \begin{equation} \label{eq:Bik}
        \mathbf{B}_{i_k} =
        \begin{pmatrix}
            \mathbf{H}_{i_1i}^\herm \tilde{\mathbf{U}}_{i_1} &
            \hdots &
            \mathbf{H}_{i_{k-1}i}^\herm \tilde{\mathbf{U}}_{i_{k-1}} &
            \mathbf{H}_{i_{k+1}i}^\herm \tilde{\mathbf{U}}_{i_{k+1}} &
            \hdots &
            \mathbf{H}_{i_{K_i}i}^\herm \tilde{\mathbf{U}}_{i_{K_i}} &
            \mathbf{A}_i
        \end{pmatrix}
        \in \complexnumbers^{M_i \times \left( \sum_{j \in \setI} d_j - d_{i_k} \right)}
    \end{equation}
    \begin{equation} \label{eq:Dik}
        \mathbf{D}_{i_k,n} = \begin{pmatrix} \Hiki \mathbf{v}_{i_k,1} & \cdots & \Hiki \mathbf{v}_{i_k,n-1} & \Hiki \mathbf{v}_{i_k,n+1} & \cdots & \Hiki \mathbf{v}_{i_k,d_{i_k}} & \mathbf{C}_{i_k} \end{pmatrix} \in \complexnumbers^{N_{i_k} \times \left( \sum_{j \in \setI} d_j - 1 \right)}
    \end{equation}
    \hrule
    \vspace{-6mm}
\end{figure*}

\vspace{-2ex}
\appendices
\section{Proof of Theorem~\ref{theorem:ia_solution_assumption_exists}} \label{proof:ia_solution_assumption_exists}
Let $d_i = \sum_{k \in \setK_i} d_{i_k}$ be the total number of data streams transmitted in cell $i$. For all $i \in \setI$, let $\mathbf{V}_i~\in~\complexnumbers^{M_i \times d_i}$ be the horizontal stacking of $\{ \Vik \}_{k \in \setK_i}$, i.e. $\mathbf{V}_i~=~\begin{pmatrix} \mathbf{V}_{i_1} & \mathbf{V}_{i_2} & \cdots & \mathbf{V}_{i_{K_i}} \end{pmatrix}$. Now study the relaxed system of equations given by
\begin{align}
    \Uik^\herm \Hikj \mathbf{V}_j &= \matO, \quad \forall \, j \in \left( \PiS \setminus \{ i \} \right), \label{eq:relaxed_IA_1} \\
    \rank{\Uik} &= d_{i_k}, \label{eq:relaxed_IA_2} \\
    \rank{\mathbf{V}_i} &= d_i, \label{eq:relaxed_IA_3}
\end{align}
for all $i \in \setI, k \in \setK_i$. If a solution exists to the original system of equations in \eqref{eq:IA_condition_intercell}--\eqref{eq:IA_condition_interstream}, then there must also exist a solution to the relaxed system of equations in \eqref{eq:relaxed_IA_1}--\eqref{eq:relaxed_IA_3}. Denote such a partial IIA solution as $\{ \tilde{\mathbf{U}}_{i_k}, \tilde{\mathbf{V}}_{i_k} \}_{i \in \setI, k \in \setK_i}$. We now constructively proceed by forming a full IIA solution $\{ \Uik, \Vik \}_{i \in \setI, k \in \setK_i}$ that will satisfy both the original system of equations in \eqref{eq:IA_condition_intercell}--\eqref{eq:IA_condition_interstream} as well as Assumption~\ref{ass:IA_properties}.

For all $i \in \setI$, let $\mathbf{A}_i \in \complexnumbers^{M_i \times \sum_{j \in \left( \PiS \setminus \{ i \} \right)} d_j}$ be the horizontal stacking of $\{ \mathbf{H}_{j_li}^\herm \tilde{\mathbf{U}}_{j_l} \}_{j \in \left( \PiS \setminus \{ i \} \right), l \in \setK_j}$. The intracoalition interference cancellation conditions in \eqref{eq:relaxed_IA_1} can now be written as
\begin{equation} \label{eq:Ai_Vi_zero}
    \mathbf{A}_i^\herm \tilde{\mathbf{V}}_i = \matO, \; \forall \, i \in \setI.
\end{equation}
From \eqref{eq:relaxed_IA_3}, we have that $\rank{\tilde{\mathbf{V}}_i} = d_i$ for all $i \in \setI$. Thus, in order for \eqref{eq:Ai_Vi_zero} to hold, we necessarily have that $\rank{\mathbf{A}_i}~\leq~M_i - d_i$. For MS $i_k$, we can now form a matrix $\mathbf{B}_{i_k}$ (see \eqref{eq:Bik} at the top of the next page) whose column space span the space that $\Vik$ is restricted from in order to satisfy \eqref{eq:IA_condition_intercell}--\eqref{eq:IA_condition_intracell}. Thus, since $\rank{\mathbf{B}_{i_k}} \leq M_i - d_{i_k}$, by solving
\begin{equation}
    \mathbf{B}_{i_k}^\herm \Vik = \matO, \; \forall \, i \in \setI, k \in \setK_i, 
\end{equation}
a new partial IIA solution $\{ \tilde{\mathbf{U}}_{i_k}, \Vik \}_{i \in \setI, k \in \setK_i}$ is found such that \eqref{eq:IA_condition_intercell}--\eqref{eq:IA_condition_intracell} as well as Assumption~\ref{ass:IA_properties}-\ref{ass:IA_properties:bullet:Vik_semiunitary} and \ref{ass:IA_properties}-\ref{ass:IA_properties:bullet:Vik_independence} are satisfied.

We will use the same reasoning in order to form $\{ \Uik \}_{i \in \setI, k \in \setK_i}$ satisfying Assumption~\ref{ass:IA_properties}-\ref{ass:IA_properties:bullet:Uik_semiunitary} and \ref{ass:IA_properties}-\ref{ass:IA_properties:bullet:Uik_independence}. For MS~$i_k$, let $\mathbf{C}_{i_k} \in \complexnumbers^{N_{i_k} \times \sum_{j \in \left( \PiS \setminus \{ i \} \right)} d_j}$ be the horizontal stacking of $\{ \mathbf{H}_{i_kj} \mathbf{V}_j \}_{j \in \left( \PiS \setminus \{ i \} \right), l \in \setK_j}$ and note that we necessarily have that $\rank{\mathbf{C}_{i_k}} \leq N_{i_k} - d_{i_k}$ due to the intercoalition interference alignment condition in \eqref{eq:IA_condition_intercell} satisfied by $\{ \tilde{\mathbf{U}}_{i_k}, \Vik \}_{i \in \setI, k \in \setK_i}$. For the $n$th stream of MS $i_k$, we can then form a matrix $\mathbf{D}_{i_k,n}$ (see \eqref{eq:Dik} at the top of the page) whose column space span the space that $\uikn$ is restricted from in order to satisfy \eqref{eq:IA_condition_intercell}--\eqref{eq:IA_condition_intracell} and the zero-forcing constraint in Assumption~\ref{ass:IA_properties}-\ref{ass:IA_properties:bullet:Uik_independence}. We see that $\rank{\mathbf{D}_{i_k,n}} \leq N_{i_k} - 1$, and thus by solving
\begin{equation}
    \mathbf{D}_{i_k,n}^\herm \uikn = \matO, \; \forall \, i \in \setI, k \in \setK_i,
\end{equation}
we obtain a solution $\{ \Uik, \Vik \}_{i \in \setI, k \in \setK_i}$ that satisfies \eqref{eq:IA_condition_intercell}--\eqref{eq:IA_condition_intracell} and Assumptions~\ref{ass:IA_properties}-\ref{ass:IA_properties:bullet:Uik_semiunitary} to \ref{ass:IA_properties}-\ref{ass:IA_properties:bullet:Vik_independence}. We can w.l.o.g. assume that \eqref{eq:IA_condition_interstream} holds a.s. due to the full rank of the involved filters and the genericness of the channel under our channel model. Finally, we can further w.l.o.g. assume that Assumption~\mbox{\ref{ass:IA_properties}-\ref{ass:IA_properties:bullet:joint_Uik_Vik_independence}} holds since no $\Hikj$ for any $j \in \PiSorth$ appeared in the derivations above. Thus, we have constructed a full IIA solution $\{ \Uik, \Vik \}_{i \in \setI, k \in \setK_i}$ that satisfies both Assumption~\ref{ass:IIA} and Assumption~\ref{ass:IA_properties}, which concludes the proof. \hfill \qedhere

\bibliographystyle{IEEEtran}
\bibliography{IEEEabrv,coordinated_precoding,rasmus_brandt}

\begin{thebibliography}{10}
\providecommand{\url}[1]{#1}
\csname url@samestyle\endcsname
\providecommand{\newblock}{\relax}
\providecommand{\bibinfo}[2]{#2}
\providecommand{\BIBentrySTDinterwordspacing}{\spaceskip=0pt\relax}
\providecommand{\BIBentryALTinterwordstretchfactor}{4}
\providecommand{\BIBentryALTinterwordspacing}{\spaceskip=\fontdimen2\font plus
\BIBentryALTinterwordstretchfactor\fontdimen3\font minus
  \fontdimen4\font\relax}
\providecommand{\BIBforeignlanguage}[2]{{%
\expandafter\ifx\csname l@#1\endcsname\relax
\typeout{** WARNING: IEEEtran.bst: No hyphenation pattern has been}%
\typeout{** loaded for the language `#1'. Using the pattern for}%
\typeout{** the default language instead.}%
\else
\language=\csname l@#1\endcsname
\fi
#2}}
\providecommand{\BIBdecl}{\relax}
\BIBdecl

\bibitem{Brandt2015}
R.~Brandt, R.~Mochaourab, and M.~Bengtsson, ``Interference alignment-aided base
  station clustering using coalition formation,'' in \emph{Proc. Asilomar Conf.
  Signals, Systems, Computers}, 2015.

\bibitem{OptResAllCoordMultiCellSys}
E.~Bj{\"{o}}rnson and E.~Jorswieck, ``Optimal resource allocation in
  coordinated multi-cell systems,'' \emph{Foundations and Trends in
  Communications and Information Theory}, vol.~9, no. 2-3, pp. 113--381, 2013.

\bibitem{Andrews2014}
J.~Andrews, S.~Buzzi, W.~Choi, S.~Hanly, A.~Lozano, A.~Soong, and J.~Zhang,
  ``What will {5G} be?'' \emph{{IEEE} J. Sel. Areas Commun.}, vol.~32, no.~6,
  pp. 1065--1082, Jun. 2014.

\bibitem{Cadambe2008}
V.~R. Cadambe and S.~A. Jafar, ``Interference alignment and degrees of freedom
  of the {K-user} interference channel,'' \emph{{IEEE} Trans. Inf. Theory},
  vol.~54, no.~8, pp. 3425--3441, 2008.

\bibitem{Mochaourab2015arxiv}
R.~Mochaourab, E.~Bj{\"o}rnson, and M.~Bengtsson, ``Adaptive pilot clustering
  in heterogeneous massive {MIMO} networks,'' \emph{Submitted to IEEE Trans.
  Wireless Commun.}, Jul. 2015, arXiv:1507.04869v1 [cs.IT].

\bibitem{ElAyach2012}
O.~El~Ayach and R.~Heath, ``Interference alignment with analog channel state
  feedback,'' \emph{{IEEE} Trans. Wireless Commun.}, vol.~11, no.~2, pp.
  626--636, Feb. 2012.

\bibitem{Krishnamachari2013j}
R.~Krishnamachari and M.~Varanasi, ``Interference alignment under limited
  feedback for {MIMO} interference channels,'' \emph{{IEEE} Trans. Signal
  Process.}, vol.~61, no.~15, pp. 3908--3917, Jul. 2013.

\bibitem{Lozano2013}
A.~Lozano, R.~Heath, and J.~Andrews, ``Fundamental limits of cooperation,''
  \emph{{IEEE} Trans. Inf. Theory}, vol.~59, no.~9, pp. 5213--5226, 2013.

\bibitem{Peters2012}
S.~Peters and R.~Heath, ``User partitioning for less overhead in {MIMO}
  interference channels,'' \emph{{IEEE} Trans. Wireless Commun.}, vol.~11,
  no.~2, pp. 592--603, Feb. 2012.

\bibitem{Chen2014}
S.~Chen and R.~S. Cheng, ``Clustering for interference alignment in multiuser
  interference network,'' \emph{{IEEE} Trans. Veh. Technol.}, vol.~63, no.~6,
  pp. 2613--2624, Jul. 2014.

\bibitem{Pantisano2013}
F.~Pantisano, M.~Bennis, W.~Saad, M.~Debbah, and M.~Latva-aho, ``Interference
  alignment for cooperative femtocell networks: A game-theoretic approach,''
  \emph{{IEEE} Trans. Mobile Comput.}, vol.~12, no.~11, pp. 2233--2246, 2013.

\bibitem{Saad2009a}
W.~Saad, Z.~Han, M.~Debbah, A.~Hj{\o}rungnes, and T.~Basar, ``Coalitional game
  theory for communication networks: A tutorial,'' \emph{{IEEE} Signal Process.
  Mag.}, vol.~26, no.~5, pp. 77--97, Sep. 2009.

\bibitem{Dreze1980}
J.~H. Dr{\`e}ze and J.~Greenberg, ``\BIBforeignlanguage{English}{Hedonic
  coalitions: Optimality and stability},''
  \emph{\BIBforeignlanguage{English}{Econometrica}}, vol.~48, no.~4, pp.
  987--1003, 1980.

\bibitem{Saad2012}
W.~Saad, Z.~Han, R.~Zheng, A.~Hjorungnes, T.~Basar, and V.~Poor, ``Coalitional
  games in partition form for joint spectrum sensing and access in cognitive
  radio networks,'' \emph{{IEEE} J. Sel. Topics Signal Process.}, vol.~6,
  no.~2, pp. 195--209, Apr. 2012.

\bibitem{Bogomolnaia2002}
A.~Bogomolnaia and M.~O. Jackson, ``The stability of hedonic coalition
  structures,'' \emph{Games Econ. Behav.}, vol.~38, no.~2, pp. 201--230, 2002.

\bibitem{Shi2011}
Q.~Shi, M.~Razavivayn, Z.~Luo, and C.~He, ``An iteratively weighted {MMSE}
  approach to distributed sum-utility maximization for a {MIMO} interfering
  broadcast channel,'' \emph{{IEEE} Trans. Signal Process.}, vol.~59, no.~9,
  pp. 4331--4340, 2011.

\bibitem{Cox1987}
H.~Cox, R.~M. Zeskind, and M.~M. Owen, ``Robust adaptive beamforming,''
  \emph{{IEEE} Trans. Acoust., Speech, Signal Process.}, vol.~35, no.~10, pp.
  1365--1376, 1987.

\bibitem{Tresch2009}
R.~Tresch and M.~Guillaud, ``Clustered interference alignment in large cellular
  networks,'' in \emph{Proc. IEEE Int. Symp. Personal, Indoor, Mobile Radio
  Commun. (PIMRC'09)}, 2009, pp. 1024--1028.

\bibitem{Park2015arxiv}
J.~Park, N.~Lee, and R.~W.~H. Jr., ``Cooperative base station coloring for
  pair-wise multi-cell coordination,'' \emph{arXiv:1503.01102 [cs.IT]}, 2015.

\bibitem{Hong2013}
M.~Hong, R.~Sun, H.~Baligh, and Z.-Q.~T. Luo, ``Joint base station clustering
  and beamformer design for partial coordinated transmission in heterogeneous
  networks,'' \emph{{IEEE} J. Sel. Areas Commun.}, vol.~31, no.~2, pp.
  226--240, 2013.

\bibitem{FundamentalsWirelessCommunication}
D.~Tse and P.~Viswanath, \emph{Fundamentals of Wireless Communication}.\hskip
  1em plus 0.5em minus 0.4em\relax Cambridge University Press, 2008.

\bibitem{ElAyach2012b}
O.~El~Ayach, A.~Lozano, and R.~Heath, ``On the overhead of interference
  alignment: Training, feedback, and cooperation,'' \emph{{IEEE} Trans.
  Wireless Commun.}, vol.~11, no.~11, pp. 4192--4203, Nov. 2012.

\bibitem{Yetis2010}
C.~Yetis, T.~Gou, S.~A. Jafar, and A.~Kayran, ``On feasibility of interference
  alignment in {MIMO} interference networks,'' \emph{{IEEE} Trans. Signal
  Process.}, vol.~58, no.~9, pp. 4771--4782, 2010.

\bibitem{Razaviyayn2012b}
M.~Razaviyayn, M.~Sanjabi, and Z.-Q. Luo, ``Linear transceiver design for
  interference alignment: Complexity and computation,'' \emph{{IEEE} Trans.
  Inf. Theory}, vol.~58, no.~5, pp. 2896--2910, May 2012.

\bibitem{Liu2013}
T.~Liu and C.~Yang, ``On the feasibility of linear interference alignment for
  {MIMO} interference broadcast channels with constant coefficients,''
  \emph{{IEEE} Trans. Signal Process.}, vol.~61, no.~9, pp. 2178--2191, May
  2013.

\bibitem{RandomMatrixTheory}
A.~M. Tulino and S.~Verdú, ``Random matrix theory and wireless
  communications,'' \emph{Foundations and Trends in Communications and
  Information Theory}, vol.~1, no.~1, pp. 1--182, 2004.

\bibitem{Mochaourab2015}
R.~Mochaourab, R.~Brandt, H.~Ghauch, and M.~Bengtsson, ``Overhead-aware
  distributed {CSI} selection in the {MIMO} interference channel,'' in
  \emph{Proc. EUSIPCO'15}, 2015, pp. 1043--1047.

\bibitem{AbramowitzStegun}
M.~Abramowitz and I.~Stegun, \emph{Handbook of Mathematical Functions}.\hskip
  1em plus 0.5em minus 0.4em\relax Dover Publications, 1965.

\bibitem{MATLAB2015a}
MATLAB, \emph{version 8.5.0 (R2015a)}.\hskip 1em plus 0.5em minus 0.4em\relax
  Natick, Massachusetts: The MathWorks Inc., 2015.

\bibitem{ElementsInformationTheory}
T.~M. Cover and J.~A. Thomas, \emph{Elements of Information Theory},
  2nd~ed.\hskip 1em plus 0.5em minus 0.4em\relax Wiley, 2006.

\bibitem{Ozarow1994}
L.~Ozarow, S.~Shamai, and A.~Wyner, ``Information theoretic considerations for
  cellular mobile radio,'' \emph{{IEEE} Trans. Veh. Technol.}, vol.~43, no.~2,
  pp. 359--378, May 1994.

\bibitem{Lapidoth1996}
A.~Lapidoth, ``Nearest neighbor decoding for additive {non-Gaussian} noise
  channels,'' \emph{IEEE Trans. IT}, vol.~42, no.~5, pp. 1520--1529, 1996.

\bibitem{IntroductoryCombinatorics}
R.~A. Brualdi, \emph{Introductory Combinatorics}.\hskip 1em plus 0.5em minus
  0.4em\relax Pearson, 2009.

\bibitem{Schmidt2013}
D.~Schmidt, C.~Shi, R.~Berry, M.~Honig, and W.~Utschick, ``Comparison of
  distributed beamforming algorithms for {MIMO} interference networks,''
  \emph{IEEE Trans. SP}, vol.~61, no.~13, pp. 3476--3489, 2013.

\bibitem{Christensen2008}
S.~Christensen, R.~Agarwal, E.~Carvalho, and J.~Cioffi, ``Weighted sum-rate
  maximization using weighted {MMSE} for {MIMO-BC} beamforming design,''
  \emph{IEEE Trans. WC}, vol.~7, no.~12, pp. 4792--4799, Dec. 2008.

\bibitem{Komulainen2013}
P.~Komulainen, A.~T\"{o}lli, and M.~Juntti, ``Effective {CSI} signaling and
  decentralized beam coordination in {TDD} multi-cell {MIMO} systems,''
  \emph{{IEEE} Trans. Signal Process.}, vol.~61, no.~9, pp. 2204--2218, May
  2013.

\bibitem{TR25814}
3GPP, ``{TR} 25.814, {Physical} layer aspects for evolved universal terrestrial
  radio access ({Release} 7),'' 3GPP, Tech. Rep., 2006.

\bibitem{Brandt2016csubmitted}
R.~Brandt, R.~Mochaourab, and M.~Bengtsson, ``Globally optimal base station
  clustering in interference alignment-based multicell networks,''
  \emph{Accepted in IEEE Signal Process. Letters}, 2016.

\bibitem{Jindal2010}
N.~Jindal and A.~Lozano, ``A unified treatment of optimum pilot overhead in
  multipath fading channels,'' \emph{{IEEE} Trans. Commun.}, vol.~58, no.~10,
  pp. 2939--2948, Oct. 2010.

\bibitem{Gomadam2011}
K.~Gomadam, V.~R. Cadambe, and S.~Jafar, ``A distributed numerical approach to
  interference alignment and applications to wireless intererence networks,''
  \emph{{IEEE} Trans. Inf. Theory}, vol.~57, no.~6, pp. 3309--3322, 2011.

\end{thebibliography}

\end{document}